\documentclass[12pt]{iopart}

\usepackage[a4paper, total={6.7in, 10in}]{geometry}

\expandafter\let\csname equation*\endcsname\relax
\expandafter\let\csname endequation*\endcsname\relax
\usepackage{amsmath}
\usepackage{graphicx}
\usepackage[rightcaption]{sidecap}
\usepackage{caption}
\usepackage{amssymb}

\usepackage{commath}
\usepackage{xcolor}
\definecolor{darkblue}{rgb}{0, 0, 0.5}
\usepackage[colorlinks,linkcolor=darkblue, citecolor=darkblue, urlcolor=darkblue]{hyperref}

\usepackage{comment}
\usepackage[numbers]{natbib}

\newcommand{\trap}{{\text{trap}}}

\newcommand{\inv}{{\text{inv}}}
\newcommand{\sca}{{\text{sc}}}
\newcommand{\Rideal}{R_\sca^{\text{ideal}}}
\newcommand{\elastic}{{\text{elastic}}}

\begin{document}

\title{Near-resonant light scattering by an atom in a state-dependent trap}

\author{T.~D.~Karanikolaou$^1$, R.~J.~Bettles$^1$, D.~E.~Chang$^{1,2}$}
\address{$^1$ICFO-Institut de Ciencies Fotoniques, The Barcelona Institute of Science
and Technology, 08860 Castelldefels (Barcelona), Spain}
\address{$^2$ICREA-Institució Catalana de Recerca i Estudis Avançats, 08015 Barcelona, Spain}
\ead{teresa.karanikolaou@icfo.eu}

\vspace{10pt}
\begin{indented}
\item[]January 2024
\end{indented}

\begin{abstract}
The optical properties of a fixed atom are well-known and investigated. For example, the extraordinarily large cross section of a single atom as seen by a resonant photon is essential for quantum optical applications. Mechanical effects associated with light scattering are also well-studied, forming the basis of laser cooling and trapping, for example. Despite this, there is one fundamental problem that surprisingly has not been extensively studied, yet is relevant to a number of emerging quantum optics experiments. In these experiments, the ground state of the atom experiences a tight optical trap formed by far-off-resonant light, to facilitate efficient interactions with near-resonant light. However, the excited state might experience a different potential, or even be anti-trapped. Here, we systematically analyze the effects of unequal trapping on near-resonant atom-light interactions. In particular, we identify regimes where such trapping can lead to significant excess heating, and a reduction of total and elastic scattering cross sections associated with a decreased atom-photon interaction efficiency. Understanding these effects can be valuable for optimizing quantum optics platforms where efficient atom-light interactions on resonance are desired, but achieving equal trapping is not feasible.
\end{abstract}

%
%
%
%
%

\section{Introduction}

The ability to trap and cool atoms using light~\cite{Metcalf2003} serves as a key enabling technique in modern experiments within the field of atomic, molecular, and optical physics, with applications spanning quantum simulation with atoms~\cite{blochreview,kaufman2021}, quantum information processing~\cite{broaways2010, lukin2018},
metrology~\cite{endres2019, kaufman2019, derevianko2011}, quantum optics~\cite{grangier2005, grangier2006} and ultracold chemistry~\cite{molecules2018}. A common trapping technique is based on the use of red-detuned, far-off-resonance light~(FORT traps)~\cite{Metcalf2003} to create a confining potential for atoms in their electronic ground states. For quantum optics applications, a number of recent experiments use this technique to precisely position atoms in a cavity~\cite{rempe2014,rauschenbeutel2021}, within a tightly focused beam~\cite{bianchet2022,Kurtsiefer2008}, or near a nanophotonic system~\cite{ rauschenbeutel2021,lukin2013} to maximize the coupling efficiency of these atoms with near-resonant photons~\cite{chang2018} in a particular optical mode. Arrays of atoms trapped periodically in an optical lattice have also been shown to exhibit efficient interactions with resonant light due to strong interference effects in light scattering~\cite{ bloch2020}.

For a FORT alone, the effect of the small electronic excited-state population induced by the off-resonant light on motional heating is well-known~\cite{savard1997,Tannoudji1989, Tannoudji2011}. However, for the quantum optics experiments above, the resonant weak driving on top of the FORT can induce additional excited population. Then, the motional potentials of both electronic ground and excited state become relevant for the optical interactions. The limit where the ground and excited states experience equal trapping potentials is well-studied, and corresponds to the problem of a trapped ion~\cite{zoller1992, zoller1994, Liebfried2003}. On the other hand, for neutral atoms, the potentials are only equal when the FORT lasers are fine-tuned to specific, ``magic'' wavelengths, which are not always available due to the atomic species, or the constraints of the experimental setup~\cite{katori2008}. Aside from the possibility of unequal trapping, the excited state might even be anti-trapped. The dependence of optical transitions on the motional properties of the atoms is an expected source of imperfections in various experiments~\cite{bloch2020, urunuela2022}.

In this work, we develop a quantum mechanical theory describing the interplay between near-resonant optical response and motion in such situations, focusing in particular on the limiting cases where the excited state is free or experiences an anti-trapping potential opposite in magnitude to the ground state potential. We elucidate on one hand how the total and elastic scattering cross sections of near-resonant light are modified, relative to the case of a stationary atom where the resonant cross section is known to have a value of $\sigma\sim \lambda^2$, where $\lambda$ is the resonant wavelength associated with the transition. A reduction of these cross sections directly reflects a reduction of the interaction efficiency between a single atom and photon, and is thus important to quantify for potential applications such as quantum memories or photon gates. On the other hand, we calculate the motional heating rate and the excess as compared to magic wavelength trapping. To our knowledge, heating for unequal trapping has only been previously treated based on a model of classical motion~\cite{martinez2018}, which is solved by Monte Carlo simulations. Besides employing a quantum formulation, we also show that in the limit of early times and weak resonant driving, the relevant rates can be obtained analytically and interpreted in terms of simple underlying intuition.

The rest of the paper is structured as follows. In Section \ref{sec:framework}, we provide a detailed explanation of our formalism, and carefully define the weak driving and early-time limits~(where the atom is unlikely to have scattered even a single photon) in which the problem significantly simplifies. In Section \ref{sec:scattering_rates}, we analyze the total and elastic scattering cross sections for various cases. In particular, we first briefly illustrate the application of our formalism on the known case of magic wavelength trapping, which also provides a useful comparison with other cases. We then consider the cases of a free and anti-trapped excited state. We find specifically that the effect of unequal trapping on the reduction of cross sections becomes significant when the ground state trap frequency begins to become comparable to the atomic radiative linewidth, $\omega_T/\Gamma\gtrsim 1 $. This situation might be relevant as experiments begin to more extensively explore narrow transitions for quantum optics.
In Section \ref{sec:heating}, we analyze the motional heating that arises for near-resonant scattering, and find that excess heating~(relative to standard recoil heating~\cite{Liebfried2003}) due to unequal trapping becomes significant once $\omega_T/\Gamma$ is comparable to the Lamb-Dicke parameter. Interestingly, when the electronic excited state experiences an anti-trapping potential, we also identify a qualitative change in the early-time dynamics once $\omega_T/\Gamma>0.5$. In particular, it is typically assumed that atoms arrive at a quasi-steady state under weak driving due to the dissipative process of spontaneous emission. However, in the regime of $\omega_T/\Gamma>0.5$, this dissipation is overcome by the anti-trapping potential, leading to an exponential growth of heating as a function of time, even within the early-time limit.

\section{\label{sec:framework} Theoretical framework}
\subsection{General formalism}
\noindent
Here, we introduce a theoretical framework for calculating both the total and elastic scattering rates of photons, based on the interaction of a near-resonant, weak coherent state with a single two-level atom that experiences state-dependent potentials. Additionally, we outline a method for evaluating the associated motional heating.

Our starting point is a master equation for the density matrix $\rho$ of the atom, which includes both the internal~(electronic) and external~(motional) degrees of freedom,
\begin{equation}\label{eq:master_equation}
    \dot{\rho}=-\frac{i}{\hbar} [\hat{H}_{\text{ext}}+\hat{H}_{\text{int}}, \rho]+\mathcal{L}[\rho].
\end{equation}
In the Hamiltonian governing the external dynamics of the atom, we allow for the possibility of internal-state dependent potentials, depending on whether the atom is in the ground state $|g\rangle$ or excited state $|e\rangle$, 
\begin{equation}\label{eq:kinetic}
     \hat{H}_{\text{ext}}=\hat{H}_{\text{ext,g}}+\hat{H}_{\text{ext,e}}\;.
 \end{equation}
The exact forms of the state-dependent potentials will be specified later. The dynamics of the two-level atom interacting with a plane-wave, monochromatic field is governed by the Hamiltonian
\begin{equation} \label{eq:internal}
    \hat{H}_{\text{int}}=-\hbar\Delta|e\rangle\langle e|+\frac{\hbar\Omega_{\text{drive}}}{2}(e^{ik_0 \hat{x}}|e\rangle\langle g|+ \text{h.c}) \;,
\end{equation}
where $\Delta=\omega_L-\omega_0$ represents the difference between laser frequency and atomic resonance frequency and $\Omega_{\text{drive}}$ denotes the Rabi frequency. Here, $k_0=\omega_0/c$ is the resonant wavevector~(in practice, we are interested in near-resonant light so that the wavevector of the field $k\approx k_0$ can be approximated by its resonant value). The mechanical effect of light associated with the absorption or emission of a photon from the driving field is described by the operator $e^{ik_0 \hat{x}}$~(for simplicity, here we only consider motion along the $x$ direction). The operator introduces a momentum displacement, associated to the momentum of the photon.

The last term in the master equation~\eqref{eq:master_equation} captures the spontaneous emission of photons, including the momentum recoil kick, and is given by~\cite{Tannoudji2011}
\begin{align*}
     \mathcal{L}[\rho]=\int d\Omega\; \Phi(\theta) \left[-\frac{\hbar}{2}\hat{J}_{\theta}^{\dagger}J_{\theta}\rho-\frac{\hbar}{2}\rho  \hat{J}_{\theta}^{\dagger}J_{\theta}+ \hbar\; \hat{J}_{\theta}\rho \hat{J}_{\theta}^\dagger\right] \;,
\end{align*}
where
\begin{equation}\label{eq:jump_operator}
    \hat{J}_{\theta}=e^{-ik_0 \cos\theta \hat{x}}\sqrt{ \Gamma}|g\rangle\langle e| 
\end{equation} 
is the set of quantum jump operators, which describe the decay of an atom from $|e\rangle$ to $|g\rangle$, accompanied by the emission of a photon of momentum $\hbar k_0$ into a direction defined by the polar angle $\theta$ and azimuthal angle $\phi$, relative to the polar axis $x$. The term $e^{-ik_0 \cos \theta \hat{x}}$ describes the projection of the imparted momentum onto the $x$ direction, and we integrate over all the possible decay directions (solid angle $\Omega$). We take the weight factor to be $\Phi(\theta)=\frac{3}{16\pi}[1+\cos^2(\theta)]$, which corresponds to an optical transition with circular polarization in the $y$-$z$ plane~\cite{zoller1992, steck2014}.

One can rewrite Eq.~\eqref{eq:master_equation} in a different form
\begin{equation}\label{eq:master_2}
    \dot{\rho}=-\frac{i}{\hbar}( \hat{H}_{\text{eff}}\rho-\rho \hat{H}_{\text{eff}}^{\dagger})+\hbar \int d\Omega \,\Phi(\theta)\;   \hat{J}_{\theta} \rho \hat{J}^{\dagger}_{\theta}\;,
\end{equation}
which separates the evolution into a part dictated by an effective, non-Hermitian Hamiltonian and a so-called jump term~(last term on the right). This lends itself to an equivalent ``quantum jump'' interpretation of density matrix evolution~\cite{molmer1992}. Then, one describes the system dynamics via a wave function that evolves through a combination of a smooth, deterministic contribution under the non-Hermitian Hamiltonian,
\begin{equation}\label{eq:eom} 
    i\hbar \frac{d|\psi(t)\rangle}{dt}=\hat{H}_{\text{eff}}|\psi(t)\rangle=\left(\hat{H}_{\text{ext}}+\hat{H}_{\text{int}}-i\frac{\hbar\Gamma}{2}|e\rangle\langle e|\right)|\psi(t)\rangle\;,
\end{equation}
and stochastically applied, discontinuous quantum jumps. In principle, repeating the calculation of the dynamics over many ``trajectories'' and averaging gives a faithful representation of the density matrix. 

Two natural basis sets to describe the motion are a Fock state basis $|n\rangle$, particularly if the internal state of the atom experiences a trapping potential~(which we assume the ground state always does), or a momentum basis $\hbar k$, which is natural if the motion is free. For example, if the excited state is free, we will express the total wave function as
\begin{equation}\label{eq:state_framework}
    |\psi (t)\rangle= \sum_n c_n(t) |g,n\rangle  +\int dk\; c(k,t)|e,k\rangle\;.
\end{equation}
On the other hand, if the excited state sees the same trapping potential as the ground state, a more natural basis is
\begin{equation}\label{eq:state2_framework}
    |\psi(t)\rangle=\sum_n c_n(t)|g,n\rangle+\sum_{n'} c_{n'}(t)|e,n'\rangle.
\end{equation}
In the case of an anti-trapped excited state, there is no natural eigenbasis. While we can solve the problem in both bases, in the main text we focus on the state representation of Eq.~(\ref{eq:state2_framework}) as we show that it provides an intuition for the dynamics.

\subsection{Definitions of rates}
\noindent
Given the wave function $|\psi(t)\rangle$, we will be interested in calculating the following rates:

\paragraph{Total scattering rate:}
The total photon scattering rate is given by
 \begin{align}
      R_{\text{sc}}(t)&=  \int d\Omega \;\Phi(\theta)\langle\psi(t)|\hat{J}_\theta^{\dagger}\hat{J}_\theta|\psi(t)\rangle=\Gamma\cdot|\langle e|\psi(t)\rangle |^2\;,
      \label{eq:R_definition}
\end{align}
i.e. it is the product of the spontaneous emission rate and the total excited state population $|\langle e|\psi(t)\rangle |^2$~(summed over all motional states).

\paragraph{Elastic scattering rate:}
The elastic scattering rate is defined as the probability per unit time that the atom scatters a photon that has the same frequency as the incoming field. In the following, we will primarily be interested in the case where the atom starts in the motional ground state~(Fock state $n=0$) of the ground-state trapping potential, and in early-time dynamics such that the probability to have scattered a photon remains low~(see the next Sec.~\ref{subsec:validity} for a more detailed definition). In that case, elastic scattering implies that the atom falls back into state $n=0$ following a jump, and the expression for the elastic scattering rate simplifies to 
 \begin{align} \label{eq:R_elastic_scattering}
      R_{\elastic}&= \int d\Omega \;\Phi(\theta) \abs{ \langle g,n=0|\hat{J}_{\theta}|
      \psi_e(t)\rangle}^2\;, 
\end{align}
where $|\psi_e(t)\rangle=|e\rangle\langle e|\psi(t)\rangle$ is the component of the total wave function where the internal state is excited.

\paragraph{Rate of phonon increase:}
Inelastic scattering processes lead to an increase of phonons per unit time~(motional heating). Under the same assumptions as above for the elastic scattering rate, the increase of phonons per unit time is given by
\begin{equation}\label{eq:phonon_increase_rate}
   \frac{ \langle\Delta n(t)\rangle}{dt}=\int d\Omega\; \Phi(\theta) \,\sum_n\; n| \langle g, n| \hat{J}_\theta|
      \psi_e(t)\rangle|^2.
\end{equation}

\subsection{Quantum jumps and validity intervals}\label{subsec:validity}
\noindent
Although the quantum jump formalism is always valid and implementable numerically, here our goal is to introduce a set of conditions in which the dynamics significantly simplify, allowing for~(mostly) analytical solutions and greatly facilitating intuition into the problem of near-resonant light scattering with unequal trapping.

First, we solve Eq.~\eqref{eq:eom} always assuming an initial condition of $|\psi(t=0)\rangle=|g,n=0\rangle$, i.e. the atom begins in both the internal and external ground states. We also assume that the driving is sufficiently weak that most of the population resides in its internal ground state, which implies that the rate at which the system undergoes quantum jumps, given by the scattering rate itself, $R_{\text{sc}}=\langle J^{\dagger}J\rangle = \Gamma \langle \psi_e(t)|\psi_e(t)\rangle$, is much smaller than the spontaneous emission rate $R_{\text{sc}}\ll \Gamma$. Under this assumption, there is a significant range of time scales given by $1/\Gamma \lesssim t \lesssim 1/R_{\text{sc}}$ where Eq.~(\ref{eq:eom}) should lead to a quasi-steady state solution $|\psi(t)\rangle\approx |\psi\rangle_{\rm st}$. Within this time, any transient behavior owing to the initial state has disappeared due to the decay term $\Gamma$ in Eq.~(\ref{eq:eom}), and it is also unlikely that the system has undergone a quantum jump. As contributions from quantum jumps and transient behavior contribute negligibly, during this time interval we expect that we can accurately calculate all relevant rates \textit{only} from the steady-state wave function, and moreover that these rates will be largely time-independent during this interval. This procedure is also often called ``adiabatic elimination'' in quantum optics literature. Interestingly, we will later find that this procedure is invalid in the case where we calculate the motional heating rate, when the excited state experiences a sufficiently strong anti-trapping potential. This case, and a modified approach, will be discussed more in Sec.~\ref{sec:heating_antitrapped}.

\subsection{Motionless atom}
\noindent
Here, we briefly review key results for a motionless atom interacting with weak coherent state light, which serves as a useful comparison for later results including motion. The scattering rate for a static atom follows a Lorentzian distribution as a function of detuning, 
\begin{align*}
    R_{\sca}^{\text{static}}=\frac{R_{\text{sc}}^{\text{ideal}}}{4\Delta^2/\Gamma^2+1}\;.
\end{align*}
    Here we define the maximum scattering rate, achieved on resonance, as 
\begin{equation}\label{eq:R_ideal}
   R_{\sca}^{\text{ideal}}=\frac{\Omega_{\text{drive}}^2 }{\Gamma}\;.
\end{equation}
When divided by the input photon flux one can also deduce the ideal, total scattering cross-section for resonant light,
$
    \sigma_{sc}^{\text{ideal}}= \frac{3}{2\pi}\lambda^2
$ which depends solely on the wavelength $\lambda=2\pi/k_0$ of the atomic transition. For a motionless atom in the weak driving limit, the total and elastic cross sections coincide. In general, if we calculate a modified scattering rate $R_{\text{sc}}$~(either total or elastic) in the presence of motion, we can take the ratio with the ideal value, which also provides the ratio of cross sections, i.e. $\cfrac{R_{\text{sc}}}{R_{\text{sc}}^{\text{ideal}}}=\cfrac{\sigma_{\text{sc}}}{\sigma_{\text{sc}}^{\text{ideal}}}$. It should be noted that key figures of merit in atom-light interfaces, such as the cooperativity in cavity QED or the optical depth in atomic ensembles, are directly proportional to the total scattering cross section~\cite{lukin2014}. Likewise, for some applications~(e.g., a photon-photon gate~\cite{kimble2004, duerr2022}) it is necessary that a photon scatters elastically, with a well-defined phase relative to the incoming photon. From that perspective, any reduction in scattering rates due to motion, and the corresponding reduction of the cross sections, directly translates to a degradation of system performance.

\section{Results on scattering rates}\label{sec:scattering_rates}
\noindent
In this section, we calculate and analyze the total and elastic scattering cross sections for three representative cases: where the ground and excited states are equally trapped~(magic wavelength trapping), where the excited state is free, and where the excited state is anti-trapped with a potential that is equal in magnitude but opposite in sign to the ground state potential.

\subsection{\label{sec:equal_trapping} Equally trapped atom}
\noindent
The case of equal trapping is already well-established~\cite{zoller1994}, but we briefly review it here to illustrate our formalism in a simple setting and to provide a comparison to other situations. Equal trapping naturally occurs for the case of a trapped ion~\cite{zoller1992,Gardiner_book}, or when a neutral atom is trapped in a magic wavelength trap~\cite{katori2008,katori2003}.

Following our general theoretical framework of Sec.~\ref{sec:framework}, the motional Hamiltonian of Eq.~\eqref{eq:kinetic} is independent of internal state and corresponds to that of a harmonic oscillator,
 \begin{equation}\label{eq:kinetic_ion}
 \hat{H}_{\text{ext}}=\hbar \omega_T \hat{n},
\end{equation}
where $\omega_T$ is the trap frequency and $\hat{n}$ the phonon number operator. The general time-dependent state can be written in the form of Eq.~\eqref{eq:state2_framework}, with the initial condition $c_{n=0}(t=0)=1$, i.e. the atom begins in the motional and internal ground state. By substituting Eqs.~\eqref{eq:internal} and~\eqref{eq:kinetic_ion} into the equation of motion~\eqref{eq:eom}, we obtain
\begin{align}
    \label{eq:eom_excited}
    i\hbar \dot{c}_{n'}=& \left(\hbar n' \omega_T-\hbar \Delta-i\hbar \frac{\Gamma}{2}\right)c_{n'}(t)+\frac{\hbar \Omega_{\text{drive}}}{2}\left(\sum_n c_n(t)\langle e,n'| e^{ik_0 \hat{x}}|e,n\rangle\right)
\end{align}
and
\begin{align*}
    i\hbar\dot{c}_{n}=\hbar n \omega_Tc_n(t)+ \frac{\hbar \Omega_{\text{drive}}}{2}\sum_{n'} c_{n'}(t)\langle e,n| e^{-ik_0 \hat{x}}|e,n'\rangle\;.
\end{align*}
\begin{SCfigure}[1.5]
\centering
\vspace{+0.5in}
    \includegraphics[scale=0.4]{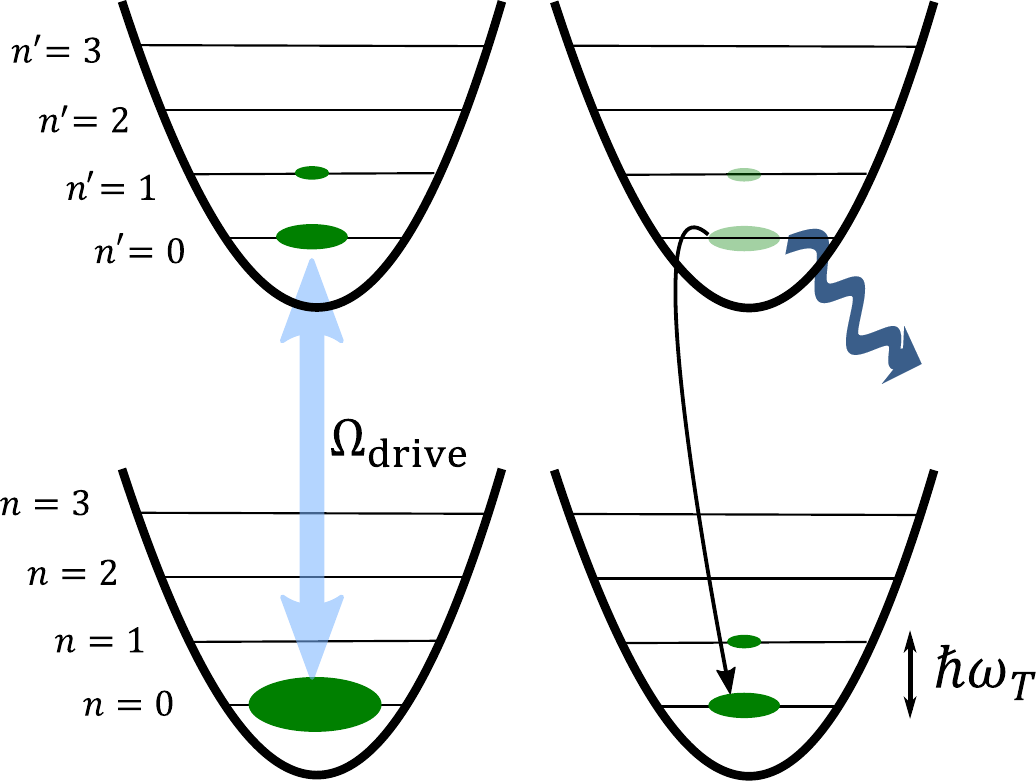}
\captionsetup{skip=1pt} 
\caption{\textbf{The excitation and decay of a magic-wavelength trapped atom in the Lamb-Dicke regime.} (Left) Initial state $|g,n=0\rangle$. A laser field $\Omega_{\rm drive}$ can generally drive the atom to a set of states $|e,n'\rangle$ characterized by phonon Fock state number $n'$. The population of mode $n'$ can be suppressed both by the Lamb-Dicke parameter $\eta$ as $\eta^{2n'}$, and by an energetic difference due to the phonon number. (Right) Upon spontaneous emission of an excited atom, the emitted photon can also cause a change in phonon number. The processes of photon absorption and emission result in atomic motional heating.}
\label{fig:ion}
\end{SCfigure}

In the weak driving and early time limits described in Sec.~\ref{subsec:validity}, the ground state population to lowest order is given by $c_n(t)=\delta_{n0}$, i.e. the population largely remains in the initial state. We can then readily solve for the~(quasi-)steady state $\dot{c}_{n'}=0$ of the excited state components, 
\begin{equation}\label{eq:ion_population}
    |c_{n'}|^2=\Omega_{\text{drive}}^2\frac{\abs{\langle n' | e^{ik_0\hat{x}} | n=0\rangle }^2}{ (2n'\,\omega_T-2\Delta)^2+\Gamma^2}\;.
\end{equation}
The distribution over different Fock states $n'$, depicted in the left panel of Fig.~\ref{fig:ion}, is generally non-trivial, due to two factors. The first, contained in the numerator, deals with the overlap between the $n=0$ ground state wave function, and the Fock states $n'$ following photon absorption. The second factor, contained in the denominator, reflects that different motional states $n'$ have different energies and thus might experience an energetic penalty to excite, given a driving field of a fixed frequency. For magic wavelength trapping, the possibility for the excited-state motional wave function to have non-trivial components $n'\neq 0$ different than that of the ground state is purely attributed to the photon momentum term, $e^{ik_0\hat{x}}$, although we will find that the situation is more complex in other cases. It is convenient to write
\begin{equation}\label{eq:exponential}
    e^{ik_0\hat{x}} =e^{i\eta(\hat{a}+\hat{a}^\dagger)}\;,
\end{equation}
where $\hat{a}$ is the phonon annihilation operator and define the \textit{Lamb-Dicke} parameter~\cite{stenholm1986}
\begin{equation}\label{eq:lamb-dicke}
    \eta =\sqrt{\frac{\omega_r}{\omega_T}}\;,
\end{equation}
with $\omega_r=\cfrac{\hbar k^2_{0}}{2m}$ being the recoil frequency. In our situations of interest, where we envision that an atom is tightly trapped, we have that $\eta\ll 1$~(known as the Lamb-Dicke limit). In this limit, the probability that the phonon number changes $|n\rangle 
\rightarrow |n'\rangle $ is suppressed as $~\eta^{2|n'-n|}$. This holds for the absorption mechanism, as well as for spontaneous emission~\cite{blatt2003} (there the Lamb-Dicke parameter appears through the jump operator of Eq.~\eqref{eq:jump_operator}).

From Eq.~\eqref{eq:R_definition} and Eq.~\eqref{eq:ion_population} in the Lamb-Dicke limit the scattering rate is
 \begin{equation}\label{eq:R_ion}
         R_\sca^{\text{equal}}=\frac{R_\sca^{\text{ideal}}}{1+\eta^2}\left[\left(\frac{4\Delta^2}{\Gamma^2}+1\right)^{-1}+\eta^2\left(\frac{(2\omega_T-2\Delta)^2}{\Gamma^2}+1 \right)^{-1}+\mathcal{O}(\eta^4)\right].
     \end{equation}
In the limit that $\Gamma\gg \omega_T$, or the so-called sideband unresolved regime, one can see from Eq.~\eqref{eq:R_ion} that on resonance ($\Delta=0$), the term in square brackets is approximately $\sim 1+\eta^2$ and thus the scattering rate is approximately equal to that of a motionless atom, $R^{\text{ideal}}_\sca$~\cite{zoller1994}. We will later see that this intuitive result -- that motion should not affect scattering when the atomic linewidth is sufficiently large -- carries over to other cases. 

For the sideband-resolved regime where $\Gamma\lesssim \omega_T$, the probability to drive the $|n=0\rangle \rightarrow |n'=1\rangle$ transition becomes strongly suppressed when $\Delta=0$, and as a result the total scattering rate on resonance is reduced by a factor $\sim(1+\eta^2)^{-1}$. Naturally, while this reduction is small in the Lamb-Dicke limit $\eta\ll 1$ for magic wavelength trapping~\cite{wineland1979}, we later find that the effects of motion can become much more pronounced when the excited state is not equally trapped, even for $\eta\ll 1$. 

From Eq.~\eqref{eq:R_elastic_scattering} the elastic scattering rate in the Lamb-Dicke regime and for $\Gamma\gg \omega_T$ is slightly reduced and given by
\begin{align}\label{eq:elastic_equal}
R_{\elastic}^{\text{equal}}= R_{\sca}^{\text{equal}}(1-\frac{7}{5}\eta^2)\;.
\end{align}
Here, a contribution of $\eta^2$ comes from the phonon added by the driving, while a factor of $\frac{2}{5} \eta^2$ comes from the photon emission~\cite{zoller1992, stenholm1986}, as schematically depicted in the right panel of Fig.~\ref{fig:ion}. 

\subsection{Atom with free excited state}\label{sec:free_excited}
\noindent
In this section, we will consider the scenario where the atomic motion is free when it is in the excited state $|e\rangle$. Following the approach outlined in Sec.~\ref{sec:framework}, we can express the motional Hamiltonian as
\begin{equation}\label{eq:h_kin_free}
    \hat{H}_{\text{ext}}=\hbar \hat{n}\omega_T |g\rangle\langle g|+ \frac{ \hbar^2 \hat{k}^2}{2m} |e\rangle\langle e|\;.
\end{equation}
The equations of motion~\eqref{eq:eom} for the evolution under the effective Hamiltonian~\eqref{eq:internal} and \eqref{eq:h_kin_free} are
\begin{align}
    \label{eq:eom_excited}
    i\hbar \dot{c}_e(k)= &\left(\frac{\hbar k^2}{2m}-\hbar \Delta-i\hbar \frac{\Gamma}{2}\right)c_e(k,t)\nonumber +\frac{\hbar \Omega_{\text{drive}}}{2}\left(\sum_n c_n(t)\langle e,k| e^{ik_0 \hat{x}}|e,n\rangle\right)
\end{align}
and
\begin{equation}
    i\hbar\dot{c}_{n}=\hbar n \omega_Tc_n(t)+ \frac{\hbar \Omega_{\text{drive}}}{2}\left(\int dk\; c_e(k)\langle e,n| e^{-ik_0 \hat{x}}|e,k\rangle\right)\;.
\end{equation}
Again, under the assumption that $c_{n}(t)\approx\delta_{n0}$ we eliminate the second equation and get for the steady state 
\begin{equation}\label{eq:steady_state_excited_free}
    c_e(k)=-\Omega_{\text{drive}}\frac{\langle k | e^{ik_0\hat{x}} | n=0\rangle }{\left(\hbar k^2/m-2\Delta\right)-i\Gamma}\;.
\end{equation}
The numerator, giving the matrix element to drive from the motional ground state to state $|k\rangle$, corresponds to a shifted version of the ground-state wave function of a harmonic oscillator in momentum space, 
\begin{equation}\label{eq:numerator}
    \langle k | e^{ik_0\hat{x}}| n=0\rangle= \left(\frac{\hbar}{m\omega_T\pi}\right)^{1/4}\cdot e^{-\frac{\hbar (k-k_0)^2}{2m\omega_T}}\;.
\end{equation}
The denominator has a $k$-dependence that reflects an energetic penalty to excite a given momentum state when its kinetic energy $\hbar k^2/2m$ is significantly mismatched from the laser frequency, specifically by an amount much larger than the natural linewidth $\Gamma$.

Knowing the wave function $c_e(k)$ allows us to directly calculate the total excited-state population and the scattering rate, as defined in Eq.~\eqref{eq:R_definition}:
\begin{equation}\label{eq:R_sc_free}
    R_{\text{sc}}=\Gamma   \int dk \;\abs{c_e(k)}^2=R_{\text{sc}}^{\text{ideal}}\int_{-\infty}^{\infty} \frac{d\tilde{k}}{\sqrt{\pi}}  \frac{e^{-(\tilde{k}-\sqrt{2}\eta)^2} }{(\frac{\omega_T }{\Gamma}\tilde{k}^2-\frac{2\Delta}{\Gamma})^2+1 }\;, 
\end{equation}
where for convenience we have normalized the wavenumber by its spread in the motional ground state, $\tilde{k}=\sqrt{\frac{\hbar}{m\omega_T}}k$. For a reasonably strong trap we get $\eta\ll \tilde{k}$, which allows us to ignore the contribution coming from the Lamb-Dicke parameter~(physically, the effect on motion is dominated by the excited state being untrapped, rather than the momentum kick of absorbing a photon from the driving field). 
\begin{figure}
    \centering
\includegraphics[width=0.65\textwidth]{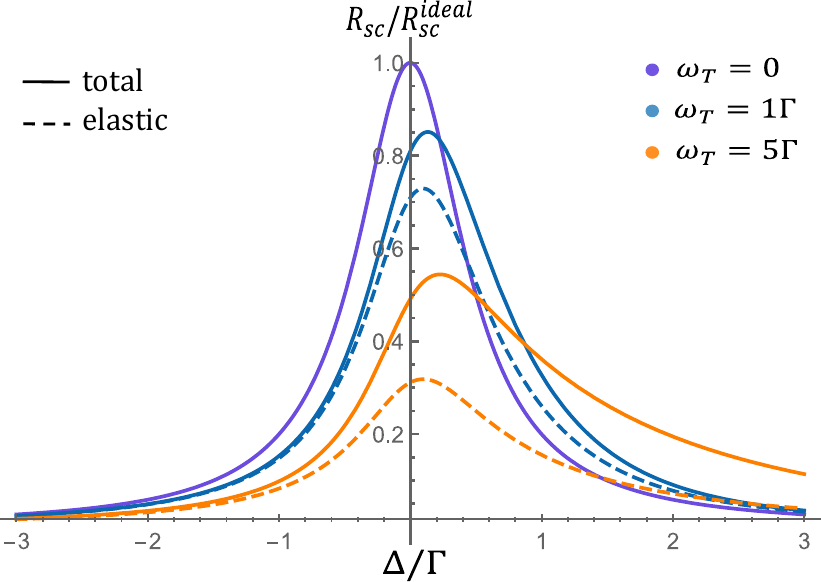}
    \caption{\textbf{Scattering rates for an atom with a trapped ground state and free excited state.} We plot the total~(solid curves) and elastic~(dashed) scattering rates as a function of dimensionless detuning $\Delta/\Gamma$, and normalize the rates by the maximal possible scattering rate $R_{\text{sc}}^{\text{ideal}}$ of a static atom on resonance. The calculations are performed for several different values of $\omega_T/\Gamma$, as indicated in the legend by different colors. For $\omega_T\ll 
    \Gamma$, one gets a Lorentzian response nearly identical to that of a motionless atom. In addition to the strong modification of scattering rates when $\omega_T\geq \Gamma$, the spectrum for the total scattering rate can develop a strong asymmetry as a function of detuning. Here we have neglected the effect of the recoil, by setting $
    \eta=0$.}
    \label{fig:R_elastic}
\end{figure}
Additionally, we find the elastic scattering rate from Eq.~\eqref{eq:R_elastic_scattering} to be
\begin{align}
    R_{\text{sc,elastic}}=\Gamma \int d\Omega\,  \Phi(\theta)\abs{\int  dk\; c_e(k)\langle n=0|e^{-ik_0 x\cos\theta}|k\rangle}^2 \;.
\end{align}
In Fig.~\ref{fig:R_elastic}, we plot both the total and elastic scattering rates~(solid and dashed curves, respectively) as functions of the detuning, assuming $\eta\approx 0$ (or equivalently $k_0 \Delta x\approx 0$) due to the arguments above. We plot these rates for several different values of $\omega_T/\Gamma$. These plots contain three prominent features: the total scattering rate on resonance is noticeably reduced once $\omega_T\gtrsim \Gamma$, the total scattering rate develops a notable asymmetry around $\Delta=0$, and the elastic scattering rate can be even further reduced but retains a more symmetric structure.

\begin{figure}
    \centering
\includegraphics[width=0.7\textwidth]{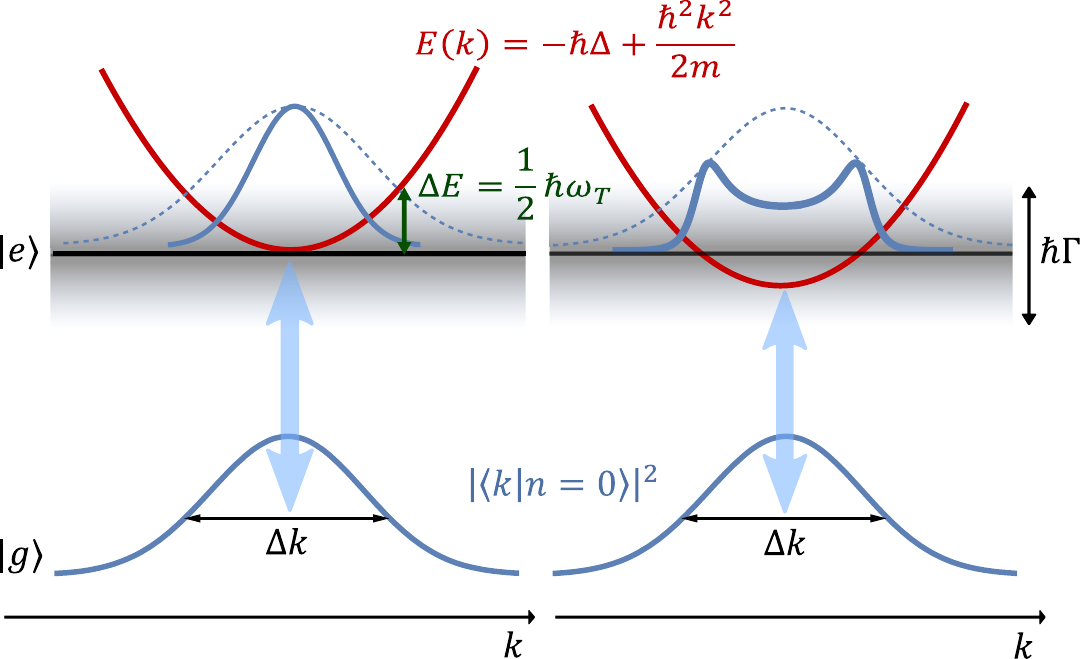}
    \caption{\textbf{Ground and excited state motional wavefunctions in $k$-space for $\Delta=0$ (left) and $\Delta>0$ (right)}. On the bottom, we plot the momentum distributions $|\psi(k)|^2=|\langle k|n=0\rangle|^2$ when the atom is in the internal and motional ground states. The distribution has a width of $\Delta k \sim \sqrt{\frac{\omega_T m}{\hbar}}$. On top, we plot the momentum distribution when the atom is in the excited state, $|c_e(k)|^2$~(solid blue curve), with the ground state distribution also drawn for reference~(dashed blue). We also plot~(red curve) the shifted dispersion relation $E(k)=-\hbar\Delta+(\hbar k)^2/2m$ of the excited state~(internal plus kinetic energy) in the rotating frame. Momentum states within an energy range $\sim\hbar\Gamma$~(gray shaded region) of the resonance condition $E(k)=0$ can be efficiently excited. In the right panel, it is evident that for positive detuning $\Delta>0$, it is possible to resonantly excite wavevectors satisfying $E(k)=0$. In contrast, for negative detuning $\Delta<0$, no such solution exists.
   }
    \label{fig:delta_asymmetry_free}
\end{figure}

The two first points can be completely understood by the kinetic energy term $\frac{\omega_T}{\Gamma} \tilde{k}^2$ in the denominator of Eq.~\eqref{eq:R_sc_free}. In these units, the ground-state wave function has a distribution of $\Delta\tilde{k}\sim 1$. However, the denominator prevents a significant fraction of this distribution of momentum states from being efficiently or resonantly excited due to energetic mismatch, once $\omega_T\gtrsim\Gamma$. This effect is illustrated first in the left panel of Fig.~\ref{fig:delta_asymmetry_free}, where we consider resonant driving $\Delta=0$. If the motional wave function associated with the excited state were to match that of the ground state~(dashed blue curve), it would have a kinetic energy spread of $\Delta E\sim \hbar\omega_T/2$. However, the natural linewidth of the excited state limits the range of wavevectors that can be efficiently excited to have an energy spread of $\Delta E\sim \Gamma$~(gray shaded region) around the resonant wavevector $k=0$, which results in a significant narrowing of the excited-state motional wave function in momentum space~(solid blue curve) and an overall reduced excited-state population~(area under the curve).

This general picture is still true when the atom is driven off resonance. However, for blue detuning ($\Delta>0$), it is possible to resonantly excite a specific wavevector satisfying $\Delta=\hbar k^2/2m$, as illustrated in the right panel of Fig.~\ref{fig:delta_asymmetry_free}. This leads to a distorted wavefunction, but which features a larger overall excited state population compared to the case of $\Delta<0$ where no wavevector components are resonant. The fact that the elastic scattering spectrum displays a higher degree of symmetry is not surprising given this same plot of $|c_e(k)|^2$. In particular, the distorted nature of the motional wave function will be largely retained after the atom emits a photon and returns to state $|g\rangle$, and it exhibits very poor overlap with the ground-state wave function of a harmonic oscillator. 

In the following, we provide analytic approximations for the total and elastic scattering rates at $\Delta=0$, while the case of $\Delta\neq 0$ can be found in~\ref{sec:scattering_appendix}. In the regime of $\omega_T\ll \Gamma$ it is straightforward to Taylor expand around small $\omega_T$ and arrive at
\begin{equation}\label{eq:analytical_omegasmall}
R_{\text{sc}}\simeq R_{\text{sc}}^{\text{ideal}}\left[1-\frac{3}{4}\left(\frac{\omega_T}{\Gamma}\right)^2\right]\;,
 \end{equation}
 and \begin{equation}\label{eq:R_el_small_approx}
   R_{\text{elastic}}\simeq \Rideal \left[1-\frac{5}{4}\left(\frac{\omega_T}{\Gamma}\right)^2\right]\;.
 \end{equation}
 For $\omega_T\gg \Gamma$ we can approximate \begin{equation}
R_{\text{sc}}=R_{\text{sc}}^{\text{ideal}} \frac{\sqrt{\pi\Gamma}}{\sqrt{2\omega_T}}\;.\label{eq:analytical_omegalarge}
 \end{equation} 
and
\begin{equation}
\label{eq:analytic_R_free_elastic} R_\elastic\simeq \Rideal \frac{\pi\Gamma}{\omega_T} \left(1-2 \sqrt{2\pi } \frac{\Gamma^{1/2}}{\omega_T^{1/2}}+4\frac{ \Gamma}{\omega_T}+\ldots\right)
\end{equation}
We can also numerically evaluate the total and elastic scattering rates as a function of $\omega_T/\Gamma$ on resonance, which we plot in Fig.~\ref{fig:scattering_antitrapped_scaling}~(blue solid and dashed curves, respectively). The asymptotic scalings agree with our derivations above. 

\subsection{\label{sec:antitrapped}Atom with antitrapped excited state}
\noindent
We now move to the case where the excited-state potential is anti-trapping. We begin by formulating a more general way of solving the equation of motion, Eq.~\eqref{eq:eom}. This formulation will be especially helpful considering that an anti-trapping potential does not have its own eigenstate basis, and that directly writing down the equations of motion in some other basis set, such as Eqs.~(\ref{eq:state_framework}) or~(\ref{eq:state2_framework}), does not obviously reveal some straightforward way to arrive at a solution. We start from the formal equation of motion for the excited-state manifold,
\begin{equation}\label{eq:eom_antitrapped}
    i\hbar \frac{d
   |\psi_e(t)\rangle}{dt}=\left(\hat{H}_{\text{ext,e}}-\hbar\Delta-i\frac{\hbar \Gamma}{2}\right)
  |\psi_e(t)\rangle+
  \frac{\hbar\Omega_{\text{drive}}}{2}|n=0\rangle \;,
\end{equation}
where $|\psi_e(t)\rangle =\langle e|\psi(t)\rangle$. Note that from now on we neglect the exponential $e^{ik_0\hat{x}}$ corresponding to the kick from the photon, as its effect is secondary in the Lamb-Dicke regime, especially when the excited state is anti-trapped. This considerably simplifies the analytical results.

The external Hamiltonian associated with the excited state is given by \begin{equation}\label{eq:h_kin_antitrapped}
    \hat{H}_{\text{ext,e}}=\frac{ \hat{p}^2}{2m}-\frac{1}{2}m\Omega_{inv}^2 \hat{x}^2\;.
\end{equation} 
Given that this Hamiltonian does not have its own eigenstate basis, as an alternative strategy we employ a Green's function formalism, writing the solution to Eq.~\eqref{eq:eom_antitrapped} as, 
\begin{equation}\label{eq:steady_definition}
|\psi_e(t)\rangle=\frac{\Omega_{\text{drive}}}{2}\int_0^t dt'\;\hat{U}(t-t')e^{i\Delta (t-t')}e^{-\frac{\Gamma}{2} (t-t')}|n=0\rangle\;,
\end{equation}
where 
\begin{equation}\label{eq:U_general}
    \hat{U}(t-t')=\exp\left[-i \frac{\hat{H}_{\text{ext,e}}}{\hbar}(t-t')\right]
\end{equation}
is the time evolution corresponding to the motional Hamiltonian~\eqref{eq:h_kin_antitrapped}. While Eq.~\eqref{eq:steady_definition} is basis-independent, the goal is to find a suitable basis to readily evaluate this expression. One possibility is the plane wave $k-$basis, which is discussed in more detail in~\ref{sec:greens_k}, and which is perhaps more generally suited to treating the dynamics beyond the limits of our stated interest.  
\begin{figure}
    \centering
\includegraphics[width=0.55\textwidth]{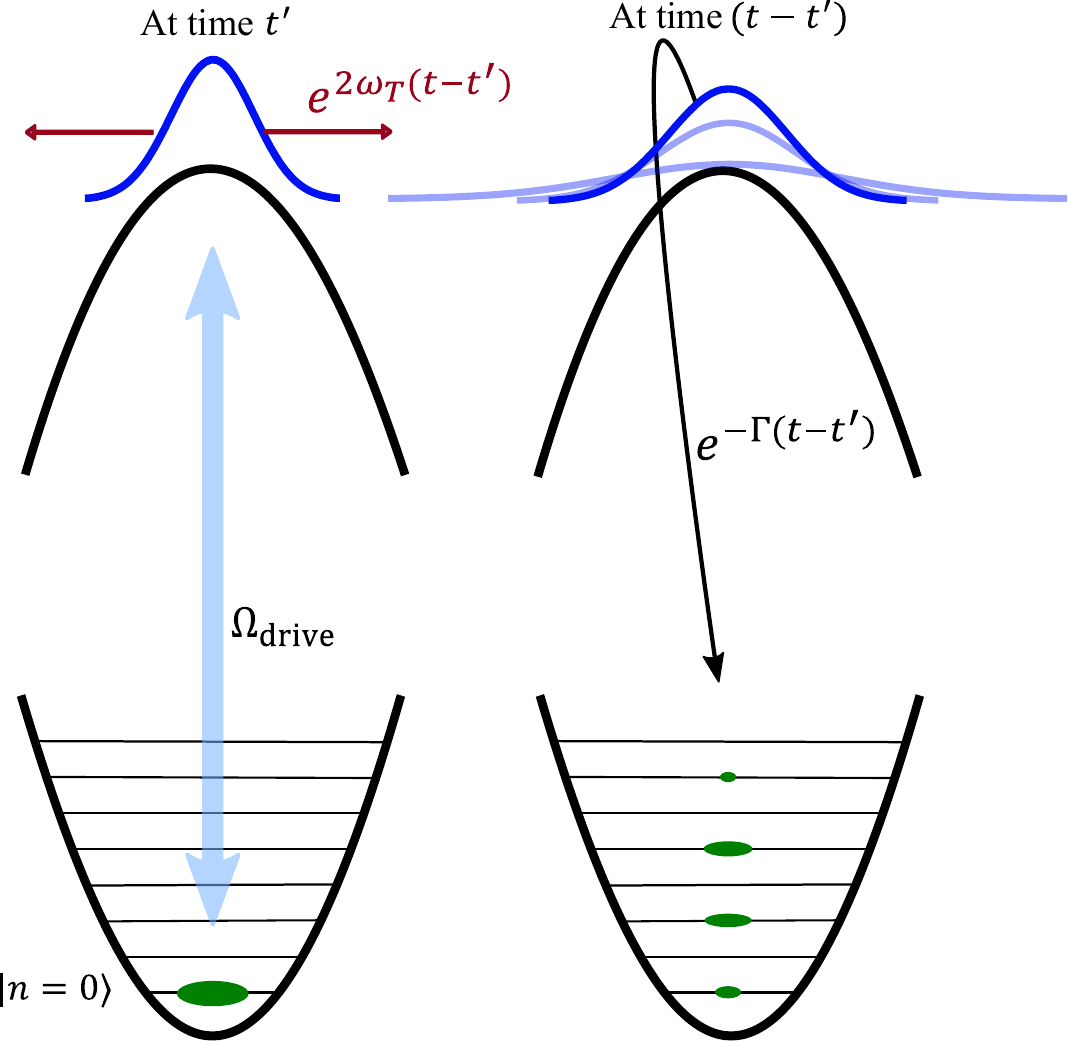}
    \caption{\textbf{Schematic representation of the time evolution given in Eq.~\eqref{eq:steady_definition}}, with initial state $|n=0\rangle$ (\textit{bottom left}). The driving allows for the excitation of a state with a rate $\Omega_{\rm drive}$ (\textit{top left}). In the excited manifold, we choose to depict the state in the more intuitive momentum basis, where $\langle k|n=0\rangle $ is a Gaussian (dark blue wavefunction). 
    This Gaussian wavefunction evolves under the influence of the inverted harmonic oscillator potential. This evolution leads to exponential expansion over time, $\langle k^2 (t-t'))\rangle\sim \exp(2\omega_T (t-t')) )$\eqref{eq:exponential_k^2}. Simultaneously, the probability of remaining in the excited state decreases exponentially with time, following $\exp(-\Gamma (t-t')))$~\eqref{eq:steady_definition}.
    The state $|\psi_e (t) \rangle$ is a superposition of Gaussian states evolved over different time intervals $\tau=t-t'$ (\textit{top right}). The resulting state after scattering is obtained by projecting  $|\psi_e(t)\rangle$ onto the ground state manifold (\textit{bottom right}). Our mathematical treatment of the problem, employing the squeezing operator from Eq.~\eqref{eq:squeezing}, allows us to describe the entire evolution purely in the Fock basis, ultimately yielding the final populations $|c_{2n}(t)|^2$ as given in Eq.~\eqref{eq:steady_inverted}. }
    \label{fig:antitrapped}
\end{figure}

Here, in the main text, we present a more elegant solution for our purposes, by working in the basis of the harmonic oscillator. We begin by expressing $\hat{x},\hat{p}$ in terms of the Fock state creation and annihilation operators associated with a \textit{normal} trap rather than an inverted one, 
\begin{align*}
       \hat{x}=\sqrt{\frac{\hbar}{2m\Omega_{\inv}}}(\hat{a}+\hat{a}^\dagger)\;,\\
        \hat{p}=\sqrt{\frac{m\Omega_{\inv}\hbar}{2}}(\hat{a}-\hat{a}^\dagger)\;,
 \end{align*}
 and thus get
\begin{equation}
    \hat{H}_{\text{ext,e}}=-\frac{\hbar\Omega_{\text{inv}}}{2}(\hat{a}^\dagger \hat{a}^\dagger+\hat{a} \hat{a})\;.
\end{equation}
The time evolution operator of Eq.~\eqref{eq:U_general} for an interval $\tau=t-t'$ is then
\begin{equation}\label{eq:squeezing}
  \hat{U}(\tau)= \exp\left[i\frac{\Omega_{\text{inv} }\tau}{2}(\hat{a}^\dagger \hat{a}^\dagger+\hat{a} \hat{a})  \right]\;.
\end{equation}

Here we will only analyze the case where $\Omega_\inv=\omega_T$ (the more general case gives no extra intuition and is better computed with the plane wave basis, see~\ref{sec:greens_k}). In that case the Fock state bases for the ground and excited state manifolds are identical. Applying $\hat{U}(\tau)$ onto a coherent state like the initial vacuum state $|n=0\rangle$ gives a time-evolved squeezed state~\cite{gaussian_information}
\begin{equation}\label{eq:evolution_inverted}
    \hat{U}(\tau)|n=0\rangle= \sum_{n=0}^\infty \frac{(-i \tanh(\omega_T  \tau))^n}{\sqrt{\cosh(\omega_T \tau)}} \frac{\sqrt{(2n)!}}{2^n n!}|2n\rangle\;.
\end{equation}
We can express the general state of Eq.~(\ref{eq:steady_definition}) in the Fock basis, $|\psi_e(t)\rangle=\sum_{n'} c_{2n'}(t)|n'\rangle$, with state amplitudes
\begin{equation}\label{eq:steady_inverted}
    c_{2n'}(t)=\langle 2n'| \int_0^{t} d\tau \; \frac{\Omega_{\text{drive}}}{2} e^{i\Delta \tau-\frac{\Gamma}{2} \tau}\hat{U}(\tau )|n=0\rangle\;.
\end{equation}

The essence of this equation is depicted in Fig.~\ref{fig:antitrapped}. An initial state $|n=0\rangle$ can become excited at arbitrary moments in time $t'$ with a rate $\Omega_{\rm drive}$. It then evolves over a time interval $\tau=t-t'$. In Eq.~\eqref{eq:steady_inverted}, we choose to describe the evolution in the Fock basis, but in the Figure we choose to depict it in the more intuitive momentum basis. The overall state at some time $t$ forms as a superposition of all possible intervals of evolution.

Within the evolution time $\tau$, there are two separate physical mechanisms at play. First is the expansion of the wave function due to evolution $\hat{U}(\tau)$ in the inverted potential. The second is the overall decay $\sim e^{-\Gamma\tau/2}$ of the excited state amplitude due to emission. When the decay dominates, the wave function decays faster than it spreads, and contributions to the wave function from time intervals $\tau\gg 1/\Gamma$ are negligible. This also sets the time it takes for the final state to reach a steady value to the order of $1/\Gamma$. Conversely, when the decay rate is small compared to the expansion rate, specifically, when $\omega_T>0.5\Gamma$, there might not be a steady-state distribution of the $c_{2n}$ or its moments, even in the weak driving limit. 

It turns out that the total population $\sum_{n}|c_{2n}(t)|^2$ does reach a steady-state value regardless of the value of $\omega_T/\Gamma$, which allows one to define a steady-state scattering rate. However, the moment $\sum_{n} 2n|c_{2n}(t)|^2$ fails to reach a steady state, which will significantly alter the calculation of heating rates. These statements will be proven in Sec.~\ref{sec:c_n_antitrapped}.
    \begin{figure}
    \centering
    \includegraphics[width=0.7\textwidth]{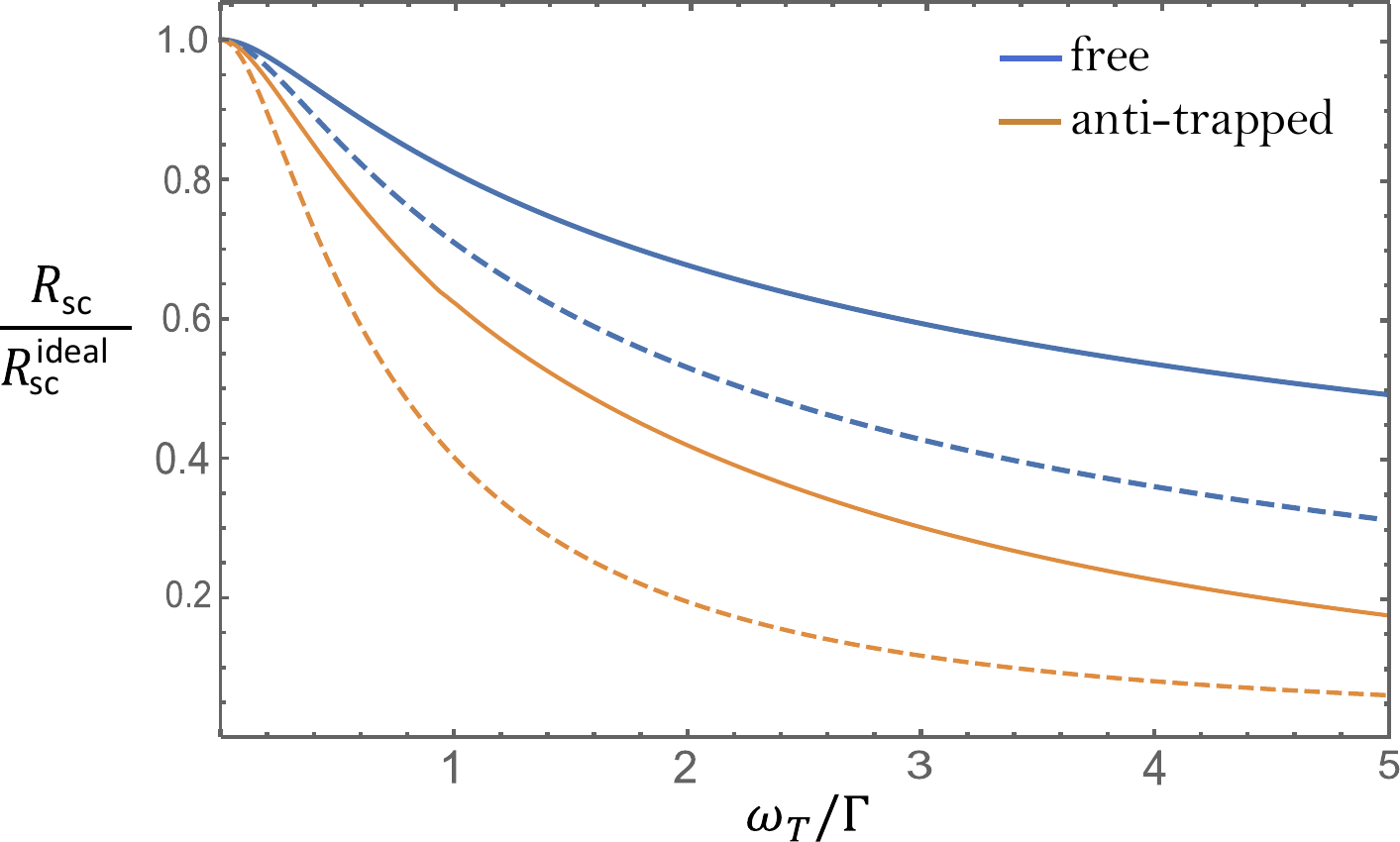}
    \caption{\textbf{Scattering rates for the free and anti-trapped potentials as a function of $\omega_T/\Gamma$, on resonance ($\Delta=0$).} The dashed lines represent the elastic scattering rates, while the solid lines represent the total rates. From top to bottom, we have found analytically and confirmed numerically that the functions scale for large $\omega_T/\Gamma$ as follows: $
\left(\frac{\omega_T}{\Gamma}\right)^{-1/2} $,  $
\left(\frac{\omega_T}{\Gamma}\right)^{-1}$,  $
\left(\frac{\omega_T}{\Gamma}\right)^{-1}$ and 
$\left(\frac{\omega_T}{\Gamma}\right)^{-2}$. The scalings are predicted in Eq.~\eqref{eq:analytical_omegalarge},  Eq.~\eqref{eq:analytic_R_free_elastic}, Eq.~\eqref{eq:analytic_R_anti} and Eq.~\eqref{eq:analytic_R_anti_elastic}.}
\label{fig:scattering_antitrapped_scaling}
\end{figure}

For now though, assuming these statements are indeed true, we can proceed to calculate the steady-state values of the scattering rates, by evaluating Eq.~\eqref{eq:evolution_inverted} at any sufficiently large value of time~(we choose $t=13/\Gamma$ in the numerics). In Fig.~\ref{fig:scattering_antitrapped_scaling}, we present the results of the numerical evaluation of the total and elastic scattering rates (depicted by orange solid and dashed curves, respectively) at resonance ($\Delta=0$). We determine these rates using their respective definitions from Eq.~\eqref{eq:R_definition} and \eqref{eq:R_elastic_scattering}. For moderate values of the trap frequency $\omega_T/\Gamma\gtrsim 1$, we observe that the scattering rates are dramatically reduced due to the anti-trapping. We next come up with analytical results for their asymptotic scalings.

\subsubsection{The population of the motional ground state \texorpdfstring{$\abs{c_0}^2$}{ }}
\noindent
\\
The population of the phononic ground state is important, as it directly tells us which fraction of the population scatters elastically. To get $\abs{c_0}^2$ we simply substitute Equation \eqref{eq:evolution_inverted} into Equation \eqref{eq:steady_inverted} and set $n'=0$.

For $\omega_T/\Gamma\ll 1$ the leading terms in the integral satisfy $\omega_T \tau\ll 1$ such that
we can approximate  \begin{equation}
       c_0\simeq \frac{\Omega_{\text{drive}}}{2}\int_0^t d\tau\; e^{i\Delta \tau-\frac{\Gamma}{2} \tau}\left(1-\frac{(\omega_T \tau )^2}{4} \right)\;,
    \end{equation}
  which in the case of $\Delta=0$ gives
    \begin{equation}\label{eq:population0_approx}
        |c_0|^2=\frac{\Omega_\text{drive}^2}{\Gamma^2} \left(1-4\frac{\omega_T^2}{\Gamma^2}+\ldots\right)\;.
    \end{equation}
 For $\omega_T/\Gamma\gg 1$, we can approximate $1/\sqrt{\cosh(\omega_T t)}\rightarrow \sqrt{2}e^{-\frac{1}{2}\omega_T t}$ . Now the integral of $|c_0|^2$ contains two exponentially decaying terms: $e^{-\frac{1}{2}\Gamma t},\,e^{-\frac{1}{2}\omega_T t}$. The first corresponds to loss of population via spontaneous emission to state $|g\rangle$, while the second represents the excitation of motion to higher Fock states $n'>0$ due to the anti-trapping potential. 
Thus the result is proportional to a Lorentzian with an enhanced linewidth, $\Gamma+\omega_T$, 
\begin{equation}\label{eq:analytic_R_anti_elastic}
\abs{c_0}^2=\frac{\Omega^2_{\text{drive}}}{\Gamma^2} \frac{2}{4\tfrac{\Delta^2}{\Gamma^2}+(\omega_T/\Gamma +1)^2}
\end{equation}
From here is is straightforward to calculate analytically the elastic scattering rate in the limits of small or large $\omega_T/\Gamma$ using Eq.~\eqref{eq:R_elastic_scattering}.

\subsubsection{Population for large n}\label{sec:c_n_antitrapped}
\noindent 
\\ In this section, we provide the scaling of $\abs{c_{2n}}^2$ as a function of $\omega_T$ and $n \gg 1$, which will lead us to the scaling of the total scattering rate and will give us a hint of how the heating rate can behave anomalously in Sec.~\ref{sec:heating}. 

We make the following set of approximations to Eq.~\eqref{eq:steady_inverted}. First, we employ Stirling's formula to approximate $\frac{\sqrt{(2n)!}}{2^n n!}\simeq (\pi n)^{-1/4}$ for large $n$. Next, we use the fact that for $n\gg 1$, the dominant contribution to the integral comes from the region of integration $\omega_T \tau>1$. This allows us to make the approximations $[\tanh(\omega_T \tau)]^n\rightarrow \exp(-2ne^{-2\omega_T \tau})$ and $\frac{1}{\sqrt{\cosh(\omega_T \tau)}}\rightarrow \sqrt{2}e^{-\frac{1}{2}\omega_T \tau}$. 
Using all three approximations we can write Eq.~\eqref{eq:steady_inverted} as 
\begin{align*}
     c_{2n}(t)\simeq\frac{\Omega_{\rm drive} \pi^{-1/4} }{\sqrt{2} } \int_0^{t} d\tau\; n^{-1/4}e^{-\frac{\omega_T+\Gamma}{2}\tau\ -2ne^{-2\omega_T \tau}}\;.
\end{align*}
We define $$2\omega_T \tau=-\ln\left(\frac{\omega_T+\Gamma}{8n\omega_T} \right)+\epsilon\;,$$ and change the variable of integration such that 
\begin{align}\label{eq:c_2n}
c_{2n}(t)\simeq&\frac{\Omega_{\rm drive} }{\sqrt{2}} \frac{\pi^{-1/4}}{2\omega_T} n^{-1/4} \left(\frac{\omega_T+\Gamma}{8n\omega_T}\right)^{\frac{\omega_T+\Gamma}{4\omega_T}} \int_{\ln\left(\frac{\omega_T+\Gamma}{8n\omega_T}\right) }^{2\omega_T t+\ln\left(\frac{\omega_T+\Gamma}{8n\omega_T}\right)} d\epsilon\; \exp\left[-\frac{\omega_T+\Gamma}{4\omega_T}\left(\epsilon+e^{-\epsilon}\right)\right]\;.
\end{align}
When discussing the steady state, we are referring to the solution where 
$t\rightarrow \infty$. This implies the upper limit of the integral becomes infinite.  At the same time
for $n\gg 1$, the lower limit of the integral approaches negative infinity. Therefore, the bounds of the integral effectively become $[-\infty, \infty]$.
Interestingly, this integral can be evaluated analytically by leveraging the definition of the Gamma-function~\cite{gammafunction},
\begin{align}
\label{eq:Gamma}
    \Gamma_G (x)&=\int_0^\infty s^{x-1}e^{-s} ds= \int_{-\infty}^{\infty} x^{x} e^{-x(\epsilon - e^\epsilon)} \, d\epsilon\;,
\end{align}
where a substitution of $s \rightarrow x\exp(-\epsilon)$ is employed.
The resulting value of the population in state $|2n\rangle $ is
\begin{equation}\label{eq:inv_approx_large_omega} 
\abs{c_{2n}}^2\simeq   \frac{1}{ \sqrt{2\pi}} \frac{\Omega^2_{\text{drive}}}{8\omega_T^2}\left( \frac{1}{2}
\right)^{\frac{\Gamma}{2\omega_T}} \frac{\Gamma_G\left(\frac{1}{4}+\frac{\Gamma}{4\omega_T}\right)^2}{n^{1+\Gamma/(2\omega_T)}} \;.
     \end{equation} 
     \begin{figure*}
    \centering
   \includegraphics[width=\textwidth]{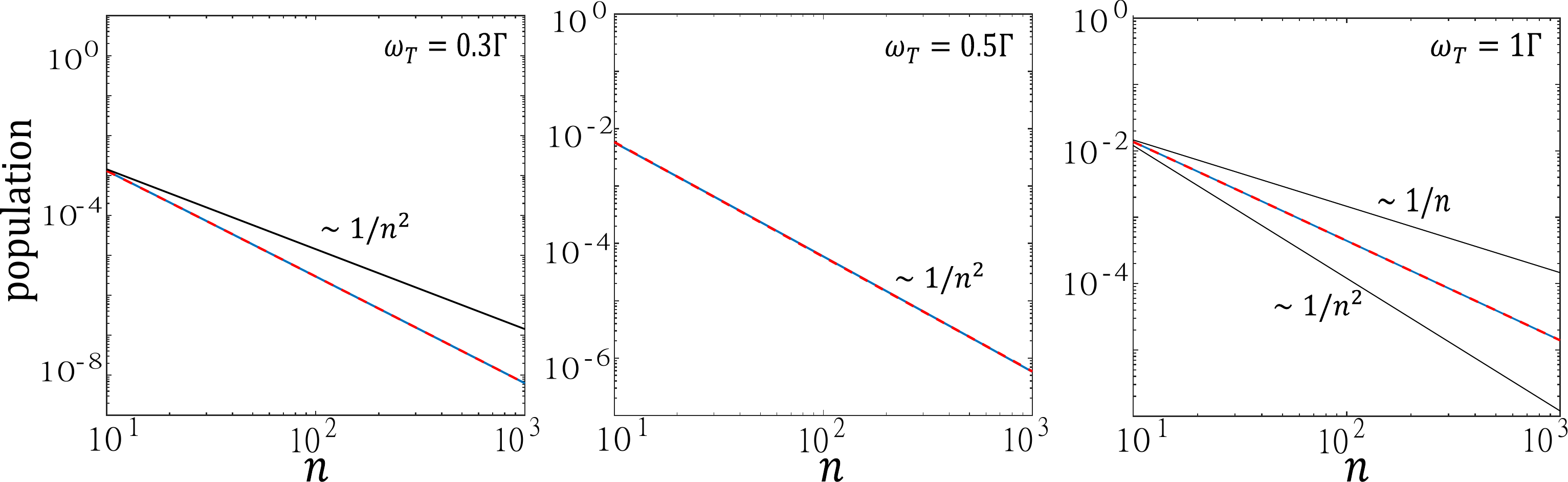}
    \caption{\textbf{Excited state population distributions for different trap frequencies.} The blue curves represent the numerically calculated populations $\abs{c_{2n}}^2$ obtained from Eq. \eqref{eq:steady_inverted} for different values of $\omega_T/\Gamma=[0.3,0.5,1]$. The red dashed line corresponds to the analytical result derived in Eq.~\eqref{eq:inv_approx_large_omega}. We observe that for large values of $n$, the approximation provided by the analytical expression is highly accurate.
  For $\omega_T/\Gamma>0.5$ the function decays faster than $1/n$ but slower than $1/n^2$. For $\omega_T/\Gamma<0.5$ it decays faster than $1/n^2$. This transition in behavior is crucial for the subsequent calculation of heating rates, which are proportional to $\sum n \abs{c_{2n}}^2$, and scattering rates, which are proportional to $\sum \abs{c_{2n}}^2$. While the first sum diverges for $\omega_T/\Gamma<0.5$, the second sum always converges.}
    \label{fig:inv_approx_large_omega}
\end{figure*}
 In Fig.~\ref{fig:inv_approx_large_omega}, we plot both the analytically approximated function of Eq.~\eqref{eq:inv_approx_large_omega} and the numerically calculated value for the population distribution $\abs{c_{2n}}^2$ as a function of $n$, for representative values of $\omega_T/\Gamma$. The results demonstrate that the approximation~\eqref{eq:inv_approx_large_omega} is remarkably accurate. 
 
Note that in all cases, $|c_{2n}|^2$ decreases with increasing $n$ faster than $1/n$, which ensures a well-defined total population (see Sec.~\ref{sec:total_anti_population}). However, only for trap frequencies $\omega_T/\Gamma>0.5$ is the decrease faster than $1/n^2$, which will play an important role in dictating the nature of early-time motional heating~(Sec.~\ref{sec:heating}).

\subsubsection{Total excited state population  }\label{sec:total_anti_population} 
\noindent
\\ We now derive approximate scalings for the total excited state population, and thereby the total scattering rate, complementing the numerical results obtained in Fig.~\ref{fig:scattering_antitrapped_scaling}.
For large $\omega_T/\Gamma$, the population distribution over Fock states is dominated by the long tails~(large $n$) of Eq.~\eqref{eq:inv_approx_large_omega}, and gives a total excited state population of 
\begin{equation} \label{eq:analytic_R_anti}
         |\langle e|\psi\rangle|^2=\sum_n^\infty  \abs{c_{2n}}^2\sim  \frac{ \Omega_{\rm drive}^2}{\omega_T \Gamma }\;.
     \end{equation}
Here, it is understood that the exact value and pre-factors can vary depending on the populations of small $n$, which are not accurately captured by the approximations leading to Eq.~(\ref{eq:inv_approx_large_omega}). Consequently, from Eq.~\eqref{eq:R_definition} the total scattering rate scales for large trap frequency as $\sim (\Gamma/\omega_T) \Rideal$. 

For $\omega_T/\Gamma\ll 1$, the value of the sum over $n$ is mainly defined by the first few $n$. Here, the portion of the range of integration that dominates Eq.~(\ref{eq:steady_inverted}) is $\omega_T \tau<1$. We can follow a procedure similar to deriving the population in the $n=0$ Fock state in Eq.~\eqref{eq:population0_approx}. For example, for $n=2$ we find
\begin{align*}
         c_{n=2}\simeq \frac{\Omega_{\text{drive}}}{2\sqrt{2}}\int_0^t d\tau e^{-\frac{\Gamma}{2} \tau}\left (1-\frac{(\omega_T \tau)^2}{4} \right)\left(\frac{i\omega_T \tau}{\sqrt{2}} \right)\;,
\end{align*}
such that after integrating we get
      \begin{equation}\label{eq:R_inv_approx}
          \sum |c_n|^2 \simeq \frac{\Omega_\text{drive}^2}{\Gamma^2} \left(1-2\frac{\omega_T^2}{\Gamma^2}\right)\;.
      \end{equation}
For $n>2$ the contributions to the sum are of higher order in $\omega_T/\Gamma$ and therefore are neglected in this first approximation.
The total scattering cross then is calculated from definition~\eqref{eq:R_definition}.

\section{Results on heating}\label{sec:heating}
\noindent
In this section, we investigate the heating that arises from light scattering. There are two distinct sources of heating to consider. The first one is the well-studied recoil heating, where the atom gains kinetic energy from the recoiled photon during the scattering process. But heating can be induced as well by the difference in potentials between the ground and the excited state of the atom.

The heating rate, which measures the energy gained per unit time, can be expressed as
\begin{equation}\label{eq:heating_rate}
     \frac{\langle 
\Delta E\rangle} {dt}= \hbar \omega_T
\frac{\langle 
\Delta n\rangle}{dt},
\end{equation}
while the early-time limit of $\frac{\langle \Delta n\rangle}{dt}$ is given in Eq.~\eqref{eq:phonon_increase_rate}~(assuming the atom starts in the ground state $n=0$). 

We begin by reproducing well-known results for an atom trapped under magic wavelength conditions~\cite{zoller1994, Tannoudji2011}. Then, the projection of the wave function onto the excited state is written as $|\psi_e\rangle=\sum_{n=0}^{\infty} c_n|n\rangle$ and the coefficients $c_n$ are given in Eq.~\eqref{eq:ion_population}. The heating rate in the Lamb-Dicke regime, which solely comes from recoil, is given by
\begin{equation}\label{eq:n_recoil}
    \frac{\langle \Delta n\rangle}{dt}\approx \frac{7}{5} \eta^2 R_{\rm sc}^{\rm equal}.
\end{equation}
The coefficient of $7/5$ directly coincides with the reduction of elastic scattering, as seen in Eq.~(\ref{eq:elastic_equal}), as expected.  Note that as the atomic dynamics reaches a steady state, the heating rate above is time-independent provided that we are in the early-time limit~(in fact, for a magic wavelength harmonic trap, the same heating rate holds for all times). This should be contrasted with the case of anti-trapping presented later. We also note that in the Lamb-Dicke regime, a tight trap~(small $\eta$) leads to a smaller rate of phonon increase. Interestingly, for unequal trapping, we will observe the opposite effect, where a tighter trap leads to an increase in heating. 

For the subsequent analysis, we will consider the case where $k_0=0$, effectively neglecting the "standard" recoil contribution and focusing solely on the excess heating resulting from unequal trapping. With this simplification the phonon-increase rate defined in Eq.~\eqref{eq:phonon_increase_rate} reduces to
\begin{equation} \label{eq:dndt} \frac{\langle \Delta n\rangle}{dt}=\Gamma \langle \psi_e(t)|\hat{n}|\psi_e(t)\rangle\;.
\end{equation}
This formulation allows us to concentrate on determining the average phonon number within the excited-state wave function.

\subsection{Heating for a free excited state}\label{sec:naverage}
\begin{figure}
       \centering
\includegraphics[width=0.7\textwidth]{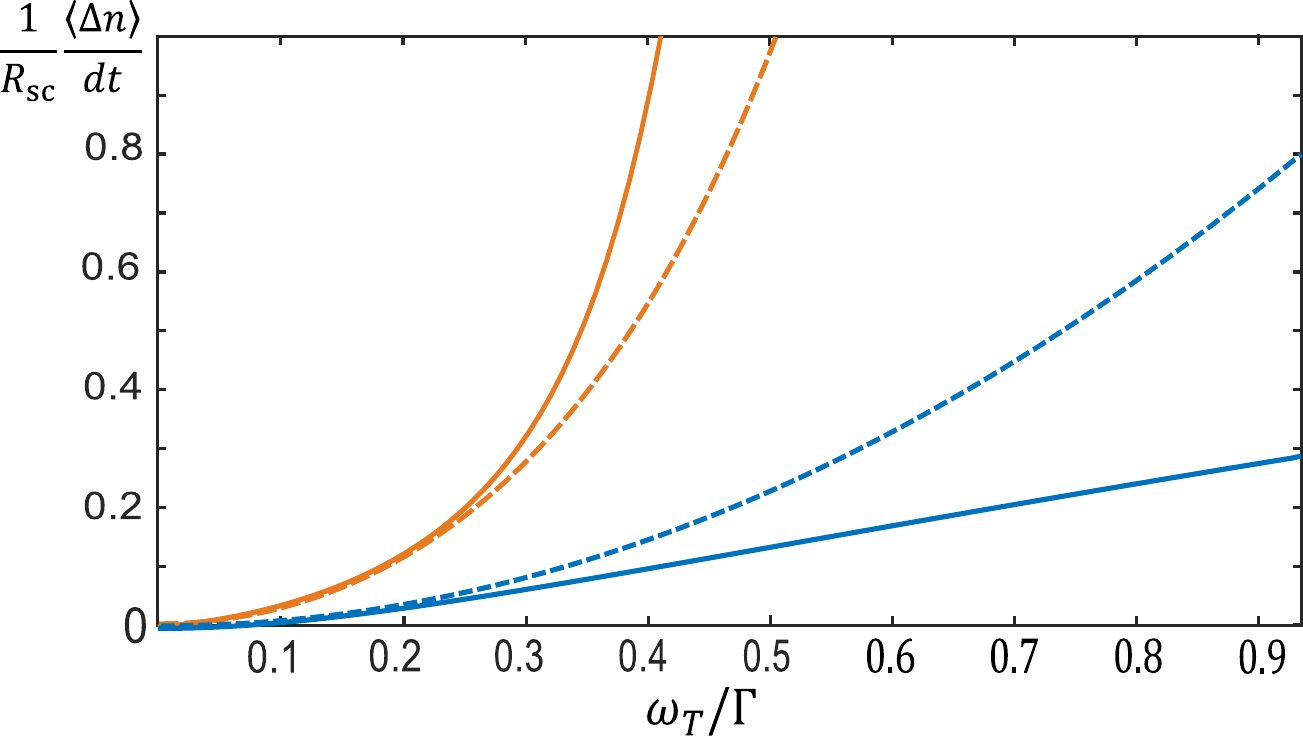}
    \caption{\textbf{The normalized heating rate $\cfrac{1}{R_{\rm sc}}\cfrac{\langle \Delta n\rangle}{dt}$ as a function of $\omega_T/\Gamma$:} (blue/orange line) the numerical results for a free/antitrapped excited state respectively. (blue/orange dashed line) Phonon number of a single $|n=0\rangle$ state which evolved for the characteristic time $t_{\rm avg}=2/\Gamma$ in a free/antitrapped potential, see Eq.~\eqref{eq:n_single_free} and \eqref{eq:n_single_anti}. }
    \label{fig:naverage_behaviour}
\end{figure}

\noindent 
In the case of a free potential, the excited state can be expressed as $|\psi_e\rangle=\int dk \; c_e(k)|k\rangle$, where the quasi-steady state values of $c_e(k)$ are given in Eq.~\eqref{eq:steady_state_excited_free}. In Fig.~\ref{fig:naverage_behaviour}, we numerically evaluate the normalized heating rate $\cfrac{1}{R_{\rm sc}}\cfrac{\langle \Delta n\rangle}{dt}$ as a function of $\omega_T/\Gamma$. We observe that for $\omega_T/\Gamma\ll 1$ and $\omega_T/\Gamma\gg 1$, the rate depends quadratically and linearly on $\omega_T/\Gamma$, respectively. The rest of this section is devoted to explaining these scalings.

\paragraph{Quadratic scaling for $\omega_T/\Gamma\ll 1$:} \noindent \\
We can estimate the heating from the simpler problem of purely free evolution of an initial Gaussian wave function~(corresponding to the $|n=0\rangle$ state), under a time $t_{\rm avg}=2/\Gamma$ corresponding to the lifetime of the excited state.  
The average energy is given by 
\begin{align*}
\left\{
\begin{aligned}
E(t_{\rm avg})&=\hbar \omega_T \left[ n(t_{\rm avg})+\frac{1}{2}\right]\\ 
 E(t_{\rm avg})&=\frac{\hbar^2  }{2m}\langle k^2(t_{\rm avg})\rangle+ \frac{1}{2}m\omega_T^2 \langle x^2(t_{\rm avg})\rangle
 \end{aligned}
 \right. \;,
\end{align*}
which connects the phonon number of the harmonic oscillator with the variances of position and momentum.

The momentum is conserved in the evolution of a free particle, $   \langle k^2(t_{\rm avg})\rangle=\langle k^2(0)\rangle$, while in real space the Gaussian wave function spreads as
 $  \langle x^2(t_{\rm avg})\rangle=\langle x^2(0)\rangle \left[ 1+\left( \omega_T t_{\rm avg} \right)^2 \right]$.
Thus we get
\begin{equation}\label{eq:n_single_free}
    n_{\rm free}(t_{\rm avg})=  \frac{1}{4}\left( \omega_T t_{\rm avg} \right)^2 \;.
\end{equation}
In Fig.~\ref{fig:naverage_behaviour} we plot the lines corresponding to $n(t_{\rm avg}=2/\Gamma)$ and see that the agreement indeed is quite good for small $\omega_T/\Gamma$. 
Thus, using the approximation $\cfrac{1}{R_{\rm sc}}\cfrac{\langle \Delta n\rangle}{dt}\sim n(t_{\rm avg})$
we conclude that for the free excited state
\begin{equation}
    \frac{1}{R_{\rm sc}}\frac{\langle \Delta n\rangle_{\text{free}}}{dt}\simeq \left(\frac{\omega_T}{\Gamma }\right)^2\;.
\end{equation} 
From this simple analysis, we can argue that the excess heating due to unequal trapping will overtake the usual contribution from recoil heating of Eq.~(\ref{eq:n_recoil}) once 
\begin{equation}\label{eq:condition_free}
\frac{\omega_T}{\Gamma} \gtrsim \sqrt{\frac{7}{5}} \eta.
\end{equation}

\paragraph{Linear scaling for \texorpdfstring{$\omega_T/\Gamma \gtrsim 0.5$:}{ }}\noindent \\
For large trap frequencies, we should return to the full expression of the excited-state wave function, Eq.~\eqref{eq:steady_state_excited_free}. An important observation is that momenta with $\cfrac{\hbar k^2}{2m}\gtrsim \Gamma $ are not efficiently excited. As a result, the wave function of the excited state possesses a momentum variance smaller than that of the ground state,  $\langle k^2\rangle_e<\langle k^2\rangle_g$, which becomes apparent for $\omega_T\gtrsim 0.5\Gamma$. In particular, from Eq.~\eqref{eq:steady_state_excited_free} we find that the variance goes as
$\langle k^2 \rangle_e\sim \frac{m\Gamma}{\hbar}$. Transitioning to real space using the Fourier limit, the maximal variance becomes $\langle x^2\rangle\sim \hbar/(m\Gamma)$. This leads to a potential energy expressed as $m\omega_T^2 \langle x^2\rangle \sim (\hbar \omega_T) (\omega_T/\Gamma)$.  Following this we get $ \cfrac{1}{R_{\rm sc}}\cfrac{\langle \Delta n\rangle_{\text{free}}}{dt}\sim (\omega_T/\Gamma) $ which explains the linear scaling observed in Fig.~\ref{fig:naverage_behaviour}.

\subsection{Heating for anti-trapped excited state}\label{sec:heating_antitrapped}
\noindent
For the antitrapped potential we write
$|\psi_e (t)\rangle=\sum_{n=0}^{\infty} c_{2n}(t)|2n\rangle$ and 
$c_{2n}(t)$ is given in Eq.~\eqref{eq:steady_inverted}. This allows us to compute the phonon-increase rate from Eq.~\eqref{eq:dndt}. 
While Eq.~(\ref{eq:dndt}) is general, we note that up to now, we have been able to replace $|\psi_e(t)\rangle$ with its steady-state value and obtain a well-defined heating rate. However, from Fig.~\ref{fig:inv_approx_large_omega}, we see that once $\omega_T/\Gamma>0.5$, the average phonon number never reaches a steady-state value, as $\langle \hat{n}\rangle$ diverges since the scaling of $|c_n|^2$ falls off slower than $1/n^2$. Below this critical trap frequency, the steady state is well defined and we plot the numerically obtained value of $\frac{1}{R_{\rm sc}}\frac{\langle \Delta n \rangle}{dt}$ in Fig.~\ref{fig:naverage_behaviour}~(solid orange curve). The divergence as $\omega_T/\Gamma\rightarrow 0.5$ can clearly be seen. We first explain the scaling for small $\omega_T/\Gamma$ immediately below, and then describe the time-dependent behavior of the heating rate when $\omega_T/\Gamma>0.5$.

\paragraph{For \texorpdfstring{$\omega_T/\Gamma\ll 0.5$:}{}} \noindent \\
As in the previous section, we use the approximation $\frac{1}{R_{\rm sc}}\frac{\langle \Delta n \rangle}{dt}\sim n(t_{\rm avg})$. For an inverse harmonic potential, both in real and momentum space, the variances spread exponentially in time. Specifically, we find that the variances in momentum and position evolve as follows:
   $$\langle k^2(t_{\rm avg})\rangle=\langle k^2(0)\rangle \cosh(2\omega_T t_{\rm avg})$$ 
   (computed in more detail in~\ref{sec:greens_k}) and 
     $$\langle x^2(t_{\rm avg})\rangle=\langle x^2(0)\rangle \cosh(2\omega_T t_{\rm avg})\;.$$
The corresponding phonon number is
 \begin{equation}\label{eq:n_single_anti}
         n_{\rm anti}(t_{\rm avg})=\frac{1}{2}(\cosh(2\omega_T t_{\rm avg})-1)\;,
     \end{equation}
and for the anti-trapped excited state (expanding the $\cosh$-function) we obtain the normalized heating rate of
\begin{equation}
     \frac{1}{R_{\rm sc}}\frac{\langle \Delta n \rangle_{\text{anti}}}{dt}\simeq 4\left(\frac{\omega_T}{\Gamma }\right)^2\;.
\end{equation} 
That explains the quadratic behaviour and allows us to state that the heating rate due to unequal trapping will be comparable to the recoil heating rate when 
\begin{equation}\label{eq:condition_inv}
\frac{\omega_T}{\Gamma} \sim \sqrt{\frac{7}{20}}\eta\;.
\end{equation}
where $\eta$ is the Lamb-Dicke parameter~\eqref{eq:lamb-dicke}. 

\paragraph{Time dependence for $\omega_T/\Gamma\geq 0.5$:}\label{sec:antitrappped_strong} 
\noindent \\
To begin with, we approximate the heating rate by an integral,
\begin{equation}\label{eq:phonon_rate_integral}
 \frac{\langle \Delta n (t) \rangle}{dt}=\Gamma \sum_{n=1}^{\infty} (2n)|c_{2n}(t) |^2 \sim 2\Gamma \int_1^{\infty} dn \;n |c_{2n}(t) |^2\;.
 \end{equation}
We can calculate the explicit time-dependence of the coefficients of $c_{2n}(t)$ obtained in Eq.~(\ref{eq:c_2n}). For that, we use the incomplete Gamma-function~\cite{gammafunction}
\begin{align*}
      \Gamma_G(x,a)=\int_a^\infty s^{x-1}e^{-s} ds=\int_{-\infty}^{-\ln(a/x)}x^{x} e^{-x(\epsilon - e^\epsilon)} \, d\epsilon\;,
\end{align*}
where the first expression is the one typically found in literature, while the second expression is derived after the change of variable $s = x\exp(-\epsilon)$. This yields the following expression
\begin{align*}\label{eq:c_t_gamma}
    c_{2n}(t) = \frac{1}{(2\pi)^{1/4}} \frac{\Omega_{drive} }{2\omega_T }   (2n)^{-\frac{1}{2}-\frac{\Gamma}{4\omega_T}} 
        \left[\Gamma_G\left(\frac{\omega_T+\Gamma}{4\omega_T}, 2n e^{-2\omega_T t}\right)-\Gamma_G\left(\frac{\omega_T+\Gamma}{4\omega_T}, 2n \right)\right]\;.
\end{align*}
We will disregard the second Gamma function, as we are primarily interested in populations of states with $n\gg 1$, and $\lim_{{n \to \infty}} \Gamma_G\left(\cfrac{\omega_T+\Gamma}{4\omega_T}, 2n \right) = 0$.
Thus, we get
\begin{equation}\label{eq:c_n_t_gamma}
  |c_{2n}(t)|^2= b \left[\Gamma_G\left(\frac{\omega_T+\Gamma}{4\omega_T}, 2n e^{-2\omega_T t}\right)\right]^2\,(2n)^{-1-\frac{\Gamma}{2\omega_T}}\;,
\end{equation}
with 
\begin{align*}
b=\frac{1}{\sqrt{2\pi}} \frac{\Omega^2_{drive} }{4\omega^2_T }\;.
\end{align*}
As a crosscheck, one can see that for $t\rightarrow \infty$,  where the Gamma-function reduces to $\Gamma_G\left(\frac{\omega_T+\Gamma}{4\omega_T}, 0\right)$, the result gives exactly the steady state population computed in Eq.~\eqref{eq:inv_approx_large_omega}. By inserting the steady-state result from Eq.~\eqref{eq:inv_approx_large_omega} into Eq.~\eqref{eq:phonon_rate_integral}, we find that for $\omega_T<0.5 \Gamma$, the phonon-increase rate grows as 
\begin{align}\label{eq:n_divergence}
    \frac{\langle \Delta n(t\rightarrow \infty)\rangle}{dt}\sim \frac{1 }{1-2\frac{\omega_T}{\Gamma }} \;. 
\end{align}
However, for $\omega_T\geq 0.5\Gamma$ the same expression gives infinity, implying that one must consider the~(small) time-dependent corrections to the steady-state result.

Therefore, for $\omega_T/\Gamma\geq0.5$, we keep the time finite. 
By inserting Eq.\eqref{eq:c_n_t_gamma} into Eq.\eqref{eq:phonon_rate_integral}, we obtain
 \begin{align*}
 \frac{\langle \Delta n \rangle}{dt}\sim  \Gamma b \int_1^{\infty}  (2 n)^{-\frac{\Gamma}{2\omega_T}}\left[\Gamma_G\left(\frac{\omega_T+\Gamma}{4\omega_T}, 2n e^{-2\omega_T t}\right)\right]^2 dn\, .
 \end{align*}
Doing the variable transformation $ 2n e^{-2\omega_T t}\rightarrow m$, 
we get 
\begin{equation}\label{eq:phonon_increase_t}
\frac{\langle \Delta n\rangle}{dt}\sim \left(\frac{\Gamma b }{2}\right) e^{(2\omega_T-\Gamma) t}  I(\omega_T,\Gamma,t)\;,
 \end{equation}
 where
 \begin{align*}
I(\omega_T,\Gamma,t)=\int_{2e^{-2\omega_T t}
    }^{\infty} \; m^{-\Gamma/(2\omega_T)} \left[ \Gamma_G\left(\frac{\omega_T+\Gamma}{4\omega_T}, m\right) \right]^2 dm\;. 
\end{align*} 
When $\omega_T t\gtrsim 0.5$, the lower limit of integration becomes less than one, allowing us to split the integral into two parts $\int_{2e^{-2\omega_T t}}^1 \ldots dm +\int_{1}^\infty \ldots dm$. 
For the first part, we can approximate it by expanding the Gamma function around small $m$, and we find that it decays as $\exp(-3\omega_T t)$. The second integral remains constant in time.
\begin{figure}
    \centering
    \includegraphics[width=0.65\textwidth]{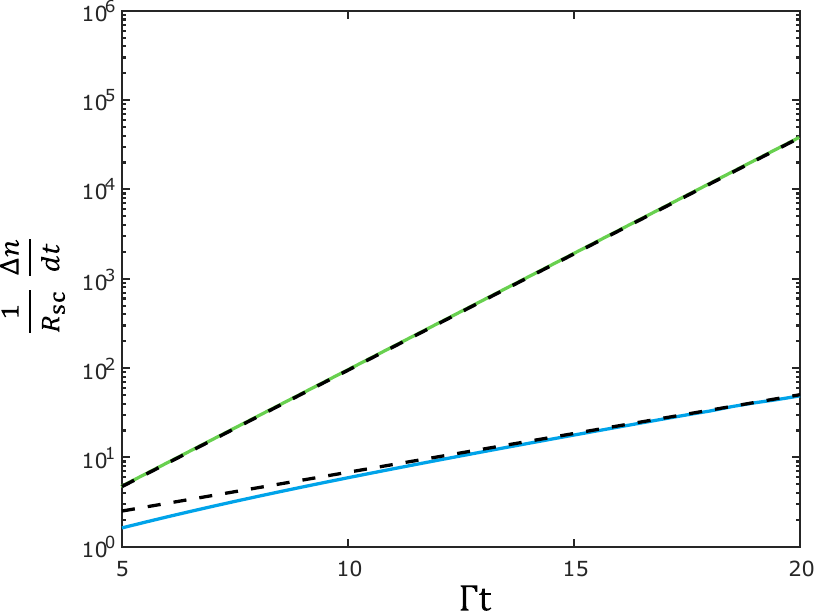}
    \caption{\textbf{The normalized heating rate  $ \frac{1}{R_{\rm sc}}\frac{\langle \Delta n(t)\rangle}{dt}$ for $\omega_T/\Gamma=0.6$ (blue) and $\omega_T/\Gamma=0.8$ (green) scale exponentially in time.} The solid lines correspond to the numerical integration of the integral Eq.~\eqref{eq:phonon_rate_integral}, where $\abs{c_{2n}}^2$ is given in Eq.~\eqref{eq:c_n_t_gamma}. 
    The dashed lines are fitted functions which scale like $\sim \exp[(2\cdot \omega_T/\Gamma -1)\Gamma t]$ for $\omega_T/\Gamma=[0.6, 0,8]$ respectively, confirming the scaling we find in Eq.~\eqref{eq:phonon_increase_t}. The approximation gets better for $\Gamma t \gg \Gamma/\omega_T$, or equivalently $t\gg 1/\omega_T$. 
    }
    \label{fig:heating_rate_t}
\end{figure}

As a result, the leading term in Eq.~\eqref{eq:phonon_increase_t} grows exponentially in time for $\omega_T t\gtrsim 0.5$. The normalized expression $ \frac{1}{R_{\rm sc}}\frac{\langle \Delta n(t)\rangle}{dt}$ has the same behavior, because the scattering rate itself becomes time-independent after a short transient time on the order of $\sim 1/\Gamma $. 
Our calculations are validated numerically in Fig.~\ref{fig:heating_rate_t}. 

To summarize, we have found that the heating rate exhibits time-dependent behavior, specifically, it increases exponentially over time. This time dependence arises from the fact that the steady state is not reached in the excited manifold. 
For a better understanding of why this time dependence takes on an exponential form, we can refer to Fig.~\ref{fig:antitrapped}. The state in the excited manifold $\psi_e(t)$ can be thought of as a superposition of Gaussian states $g(k)$, each evolved for different time intervals $\tau = t - t'$, as described in Eq.~\eqref{eq:steady_inverted}. Importantly, the variance (or equivalently, the phonon number) of each Gaussian grows exponentially as $\exp(2\omega_T \tau)$, while the probability of the state to remain in the manifold decreases exponentially as $\exp(-\Gamma \tau)$.
When the exponential spread dominates over the decay, every Gaussian contributes significantly to the heating, regardless of how long it has evolved. The maximal contribution to heating comes from the spread of the longest evolved Gaussian, which for a given time $t$ is $\exp [(2\omega_T-\Gamma)t)]$.

\section{Discussion}
\noindent
In summary, we have elucidated the near-resonant optical properties of an atom that experiences strongly state-dependent potentials. Our results show that the optical properties for an atom with a free or anti-trapped excited state can become significantly modified with increasing ratios of trap frequency to linewidth, $\omega_T/\Gamma$. For small values of $\omega_T/\Gamma\gtrsim \eta$ that exceed the Lamb-Dicke parameter, we show that the excess heating due to the unequal potentials can already become comparable to the heating rate expected due to photon recoil. For $\omega_T/\Gamma\gtrsim 1$, the interaction efficiencies of an atom with near-resonant light, as captured by the total and elastic scattering cross sections, can be significantly reduced compared to the ideal values of a static atom. We anticipate that these results can serve as a practical guide to design quantum optics experiments involving tightly trapped atoms, either when magic wavelength traps are not feasible and/or when narrow atomic transitions are used.

As a concrete example, the transition $ ^1S_0 \rightarrow ^3P_1 $ of strontium possesses a narrow linewidth of $7.5$ kHz. For a trapping frequency of $\omega_T/\Gamma\sim 5$, the interaction efficiency can be reduced to just $0.6\%$ or $0.2\%$ of the maximum value should the excited state be left free or anti-trapped, respectively. Simultaneously, early-time heating is expected to be roughly 100 times stronger than the recoil contribution for a free excited state, while for an anti-trapped state, it exhibits time-dependent and exponential growth.

While our analysis focused on the weak driving and early-time limits, in principle our theoretical approach is general, and could at least be implemented numerically beyond these regimes. Besides numerically exact results, it would be interesting in future work to develop simpler semi-classical descriptions that nonetheless contain the key physics. These semi-classical descriptions would ideally be able to capture anharmonic traps, and thus atom escape rates from realistic finite-depth traps, which cannot readily be done within our approach. It might also be interesting to investigate whether the possibility for cooling exists, based on state-dependent potentials.
These directions could provide valuable insights into the behavior of trapped atoms under different conditions and potentially open up new avenues for experimental exploration.

\section*{Acknowledgments}
\noindent
T.D.K. acknowledges support from the European Union’s Horizon 2020 research and innovation programme under the Marie Skłodowska-Curie grant agreement No 847517. R.J.B. acknowledges support from the Engineering and Physical Sciences Research Council [Grant No. EP/R002061/1]. D.E.C. acknowledges support from the European Union, under European Research Council grant agreement No 101002107 (NEWSPIN), FET-Open grant agreement No 899275 (DAALI) and EIC Pathfinder Grant No 101115420 (PANDA); the Government of Spain under Severo Ochoa Grant CEX2019-000910-S (MCIN/AEI/10.13039/501100011033); Generalitat de Catalunya (CERCA program and AGAUR Project No. 2021 SGR 01442); Fundació Cellex, and Fundació Mir-Puig.

\clearpage
\bibliographystyle{apsrev4-1}
\bibliography{references}

\clearpage
\markboth{}{}
\appendix
\section{Scattering rate for a free excited state as a function of \texorpdfstring{$\Delta$} {delta}}\label{sec:scattering_appendix}
\noindent
In Fig.~\ref{fig:R_elastic}, it is evident that the total scattering rate exhibits an asymmetry with respect to $\Delta$. When the objective is to maximize the scattering rate, having knowledge of the optimal detuning is valuable. Consequently, in the following, we provide approximate expressions for the total scattering rate of an atom with a free excited state as a function of $\omega_T$, $\Gamma$, and $\Delta$. For $\omega_T< \Gamma$ the scattering rate can be approximated by
\begin{eqnarray*}
    R_{\text{sc}}\simeq \frac{R_{\text{sc}}^{\text{ideal}}}{4\Delta^2/\Gamma^2+1}
 \left[1+ \frac{4\Delta \omega_T}{4\Delta^2+\Gamma^2} -\frac{3}{4}\frac{\omega_T^2 }{4\Delta^2+\Gamma^2}+\mathcal{O}\left(\frac{\omega_T ^3}{\Gamma^3}\right) \right]\;.
 \end{eqnarray*}
The second term of the sum introduces the observed asymmetry, and this asymmetry becomes more pronounced as $\omega_T$ increases. The value of the detuning that gives maximal scattering at first approximately grows linearly with trap frequency, $\Delta_{\rm max}\sim \omega_T/2$.

In order to obtain the analytic expression for large trap fequencies $\omega_T\gg \Gamma$, we use that the steady state function $c_e(k)$
in Eq.~\eqref{eq:R_sc_free} is a very narrow function that is centered around $\tilde{k}=0$ for $\Delta<0$, while for $\Delta>0$ it is centered around  $\tilde{k}=\sqrt{2\Delta/\omega_T}$. Thus, we replace the exponential $e^{-\tilde{k}^2}$ in the numerator of Eq.~\eqref{eq:R_sc_free} with $e^{-\tilde{k}^2}\rightarrow 1$ for $\Delta<0$ and with $e^{-\tilde{k}^2}\rightarrow e^{-2\Delta/\omega_T}$ for $\Delta>0$ (while at the same time we set $\eta=0$, as explained in the main text). 
With that, we can calculate the integrals analytically and obtain 
  \begin{align*}
\frac{R_{\text{sc}}}{R_{\text{sc}}^{\text{ideal}}}= \frac{\sqrt{\Gamma\pi}}{\sqrt{2\omega_T}}\frac{\sqrt{\Gamma}}{\sqrt{4\Delta^2+\Gamma^2}}  \sqrt{\sqrt{4\Delta^2+\Gamma^2}+2\Delta)}\;, 
\end{align*}
and 
 \begin{align*}
\frac{R_{\text{sc}}}{R_{\text{sc}}^{\text{ideal}}}=\frac{\sqrt{\Gamma\pi}}{\sqrt{2\omega_T}}\frac{e^{-2\Delta/\omega_T}\sqrt{\Gamma}}{\sqrt{4\Delta^2+\Gamma^2}}  \sqrt{\sqrt{4\Delta^2+\Gamma^2}+2\Delta)}\;.
\end{align*}
for $\Delta\leq 0$ and $\Delta>0$ respectively. 
The maximal scattering rate appears at small positive detunings (see Fig.~\ref{fig:R_elastic}), around which we expand and get
\begin{align*}
    \frac{R_{\text{sc}}}{R_{\text{sc}}^{\text{ideal}} } \simeq & \sqrt{\frac{\Gamma \pi}{2\omega_T}}(1-2\Delta/
\omega_T\nonumber +2\Delta^2/\omega_T^2) \cdot ( 1+\Delta/\Gamma-\Delta^2/\Gamma^2-2\Delta^3/\Gamma^3)\;.
\end{align*}
The optimal detuning $\Delta$ (where $R_{sc}$ is maximal) saturates at a value of $
\Delta_{\rm max}\simeq 0.27\Gamma $.

\section{Scattering rate for an antitrapped potential with \texorpdfstring{$\Delta\neq 0$}{delta0} \texorpdfstring{and $\Omega_{\rm inv}\neq \omega_T$}{ } }\label{sec:greens_k}
\noindent
When dealing with the case where the excited state is anti-trapped but has a different frequency than the ground state trap frequency $\omega_T$, the analysis becomes more complicated than the one in Sec.~\ref{sec:antitrapped}. This is because the Fock basis is linked to the frequencies, resulting in different Fock bases for the ground and excited states. As a result, either the initial state (written in the excited state basis) or the evolution operator of Eq.~\eqref{eq:evolution_inverted} (written in the ground state basis) takes on a more complex form~\cite{Qiu2022}.

Therefore, we introduce a different approach that is more suitable for the generalized case, as compared to the approach of Sec.~\ref{sec:antitrapped}. In this alternative approach, we project Eq.~\eqref{eq:eom_antitrapped} onto momentum space. 
The solution of the differential equation can then be expressed in terms of the Green's function $G(k, k',\tau)=\langle k|\hat{U}(\tau)|k'\rangle$, leading to
\begin{align}
\psi_e(k,t)=\frac{\Omega_{\text{drive}}}{2}\nonumber \int_0^t d\tau\;e^{(i\Delta-\frac{\Gamma}{2})\tau }\int dk'\;G(k,k',\tau)g(k')\;.
\end{align}
Here, $\psi_e(k,t)=\langle k| \psi_e(t)\rangle$, and the function $g(k)=\langle k|n=0\rangle$ is determined by Eq.~\eqref{eq:numerator} with $k_0=0$. The specific form of the Green's function can be obtained by solving the homogeneous part of Eq.~\eqref{eq:eom_antitrapped}.

\begin{figure}
    \centering
    \includegraphics[width=\textwidth]{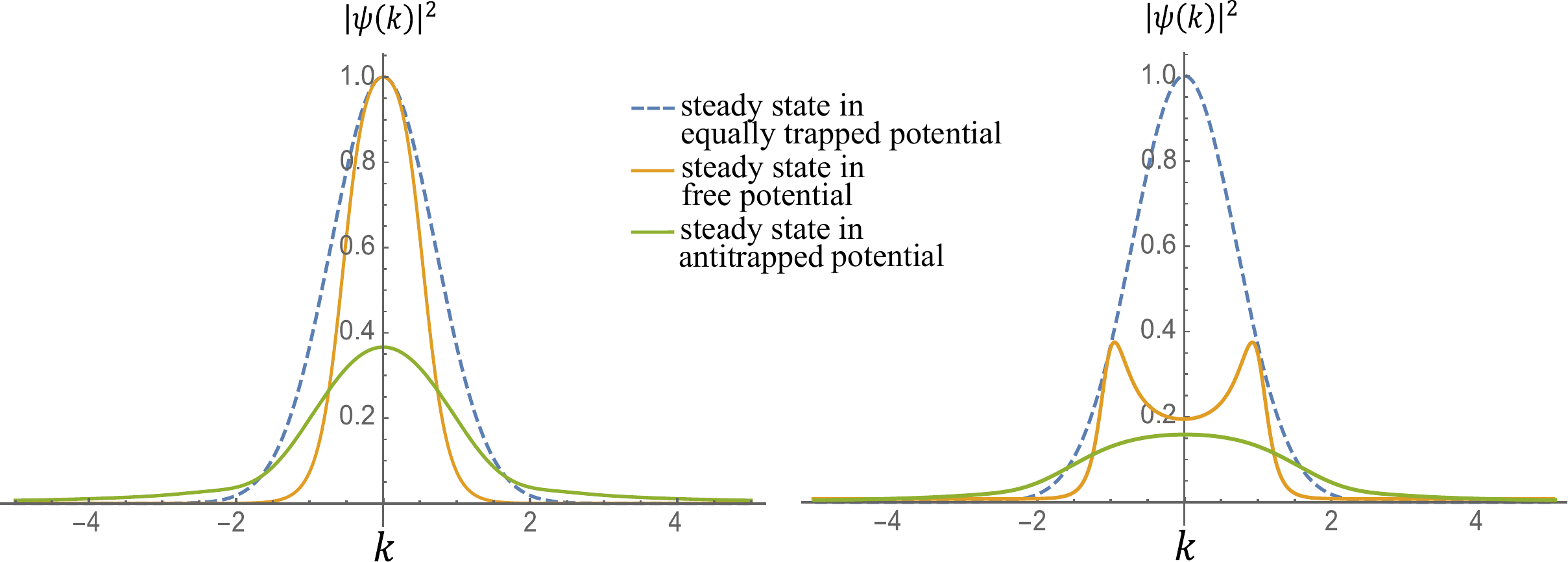}
    \caption{\textbf{Excited state momentum distributions.} Steady-state distributions of $\abs{\psi_e(k)}^2$ for a ground-state trap of frequency $\omega_T=2 \Gamma$, and the following excited state potentials: an anti-trap with frequency
    $\Omega_\inv=\Gamma$ (green), a free state (yellow) and an equal trap (dotted). The steady states for unequal trapping exhibit a reduced area compared to the equally trapped atom. On the left side we plot the resonant driving case, $\Delta=0$, while on the right side, we consider $\Delta/\Gamma=1$.}
    \label{fig:steady_state_inv_k}
\end{figure}

In the case of an anti-trapped potential, the Green's function for the evolution corresponds to the Mehler kernel~\cite{quantum_propagators} of the harmonic oscillator, with $\omega_T$ replaced by $i\Omega_\inv$. This leads to the following expression for the Green's function
\begin{align}
    \label{eq:antitapped_greens}
        G(k,k',\tau)=\sqrt{\frac{\hbar}{2m\pi}} \sqrt{\frac{\hbar}{i \sinh(\Omega_{\inv}\tau)}} \exp\left[ \frac{i\hbar kk' }{m \sinh(\Omega_{\inv}\tau)}\right] \exp\left[i \frac{\hbar(k^2+k'^2)}{2m}\coth(\Omega_{\inv}\tau) \right]\;.
\end{align}
It is important to note that a similar analysis can also be performed for a trapped or free excited state using the corresponding Green's functions~\cite{quantum_propagators}. In Fig.~\ref{fig:steady_state_inv_k}, we illustrate the momentum distribution of the excited state, $|\psi_e(k)|^2$ in steady state, for various excited-state potentials and detunings. Notably, we observe three points that make the evolution for an anti-trapped potential differ: (1) For $\Delta \neq 0$, the distribution remains peaked around $k=0$ unlike the case of a free excited state, (2) the momentum distribution is broader than the case of the magic wavelength trap~(rather than narrower, as is the case for a free excited state), and (3) the total population is reduced compared to the free potential case.

Point (1) can be naturally attributed to the fact that the anti-trapping potential has no eigenstates, which also should lead to a more symmetric scattering rate in $\Delta$. Point (2) can be explained by examining the evolution of a single Gaussian, as represented by the integral $ g(k,\tau) =\int dk' \,G(k,k',\tau)g(k')$ in Eq.~\eqref{eq:steady_definition}. The calculation yields
\begin{align*}
     g(k,\tau) =&   \mathcal{N}(\tau) \exp\left[\frac{\hbar k^2}{2m\omega_T}\frac{-i\cosh(\Omega_\inv \tau)+\frac{\omega_T}{\Omega_\inv}\sinh(\Omega_\inv \tau)}{\sinh(\Omega_\inv  \tau)\frac{\Omega_\inv}{\omega_T}+i\cosh(\Omega_\inv \tau) }\right]\;,
\end{align*}
where
\begin{align*}
    \mathcal{N}(\tau)=i\left(\frac{\hbar }{\pi
       m\omega_T}\right)^{1/4}(-i\sinh(\Omega_\inv \tau)\frac{\Omega_\inv}{\omega_T}+\cosh(\Omega_\inv \tau))^{-1/2}\;.
\end{align*}
The evolution reveals that the variance of the Gaussian evolving in an antitrapped.
potential for a time interval $\tau$ grows as \begin{equation}\label{eq:exponential_k^2}
   \langle k^2 (\tau) \rangle \sim\frac{\Omega^2_\inv+\Omega^2_\trap}{2\Omega^2_\trap} \cosh(2\Omega \tau)\rightarrow \frac{\Omega^2_\inv+\Omega^2_\trap}{4\Omega^2_\trap} \exp(2\Omega \tau)
\end{equation}  which is also what we use in Sec.~\ref{sec:naverage}. 
Thus, the exponentially increasing variance leads to stronger heating compared to the free potential (where the variance is constant). 

As for point (3), we can attribute the smaller populations to a stronger destructive interference. When integrating over all time intervals, see Eq.\eqref{eq:evolution_inverted}, the superimposed Gaussians interfere destructively, resulting in a reduced coherent driving and consequently a reduced scattering rate.
In $g(k,t)$ we see that for larger values of $k$ the phase changes are faster in time and will enhance destructive interference. This effect is stronger than in the case of a free excited state, exactly because the variance of the $k$ modes is growing in time and allows for population of large-$k$ modes.

\begin{figure}
    \centering
    \includegraphics[width=0.72\textwidth]{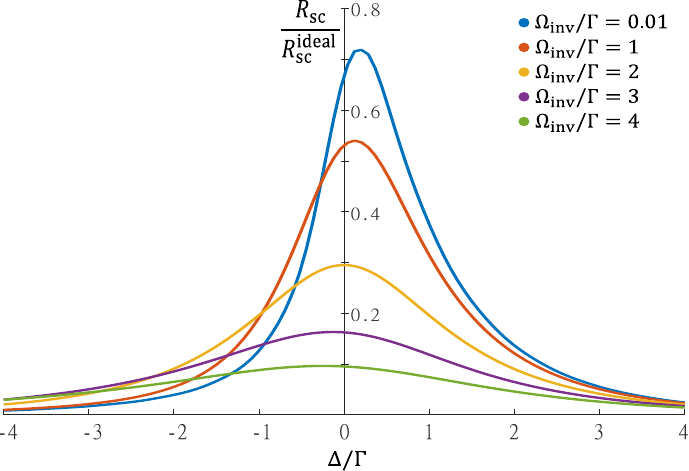}
    \caption{\textbf{Spectra of scattering rates for various anti-trapping potential strengths.} Here, we plot the normalized scattering rate $R_{\rm sc}/R_{\rm sc}^{\rm ideal}$ as a function of dimensionless detuning $\Delta/\Gamma$, for various values of $\Omega_\inv/\Gamma=[0.01, 1, 2, 3, 4]$, with $\omega_T/\Gamma=2$. At $\Omega_\inv=0.01$, the results align with those of a free excited potential, as expected. When $\Omega_\inv=\omega_T$, no asymmetry is observed, while otherwise, the direction of the asymmetry depends on which quantity, $\Omega_\inv$ or $\omega_T$, is larger.}
\label{fig:antitrapped_delta_scattering}
\end{figure}

In Fig.~\ref{fig:antitrapped_delta_scattering} we plot the scattering rate as a function of detuning, for various values of anti-trapping potential strengths $\Omega_{\inv}/\Gamma$. We see that the degree and direction of asymmetry in the spectra depend on the value of $\Omega_{\inv}/\omega_T$.

\end{document}